\documentclass[useAMS,usenatbib,referee]{mn2e}
\usepackage{graphicx}
\usepackage{natbib}
\usepackage{times}
\usepackage{float}
\usepackage{epsfig}

\usepackage{amsmath}

\title[A twisted  disc in  binary  black holes]{On the evolution of a twisted thin accretion 
disc in eccentric inclined supermassive binary black holes
 }
\author[ P. B. Ivanov and V. V. Zhuravlev ]{P.B.Ivanov$^{1}$\thanks{E-mail:
pbi20@cam.ac.uk (PBI)}, V. V. Zhuravlev $^{2}$\thanks{E-mail:
v.jouravlev@gmail.com (VVZ)}\\
$^{1}$Astro Space Centre, P.N. Lebedev Physical Institute, 84/32
Profsoyuznaya Street, Moscow, 117997, Russia  \\
$^{2}$ Sternberg Astronomical Institute, Lomonosov Moscow State University, Universitetskij pr., 13, Moscow 119234, Russia }

\begin{document}

\date{Accepted. Received; in original form}

\pagerange{\pageref{firstpage}--\pageref{lastpage}} \pubyear{2010}

\maketitle

\label{firstpage}

\begin{abstract}
We propose a model of a  twisted accretion disc around a Kerr black hole
interacting with a secondary black hole of a smaller mass on an inclined eccentric orbit.
We  use parameters of the system, which may be appropriate for 
the so-called 'precessing massive' model of OJ 287.
We calculate expressions for torque exerted on the disc by the secondary and a contribution of the secondary 
to the apsidal precession of disc elements by a double averaging procedure over the periods of
the secondary and the disc elements. These expressions are used at all
scales of interest, including the ones inside the binary orbit. We calculate numerically
the evolution of the disc tilt and twist assuming a flat initial configuration. We consider the disc
aspect ratio $h/r=10^{-3}$, a rather large viscosity parameter $\alpha=0.1$ and several values of the primary
rotational parameter, $\chi$. We find that, after a few periods of Lense-Thirring precession of the 
orbit, the disc relaxes to a quasi-stationary configuration in the precessing frame with a non-trivial 
distribution of the disc inclination angle, $\beta$, over the radial scale. 
{  We propose an analytic model for this configuration. We show that
the presence of the twisted disc leads to multiple crossings of the disc by the secondary per one orbital period, with time periods  between the  crossings being different from the flat disc model. Our results should be taken into account in the modelling of OJ 287. They can also be applied to similar sources.}  
\end{abstract}

\begin{keywords}
accretion, accretion discs, black hole physics,  hydrodynamics, galaxies:
interactions, galaxies: nuclei
\end{keywords}

\section{Introduction}

Supermassive binary black holes (SBBHs) forming as a result of the inevitable process of galactic mergers could lead to many striking observational effects, see e.g. \cite{Fer}, \cite{Kom1} and \cite{Dot} for general reviews. Typically, candidates to SBBHs are identified as quasars and active
galactic nuclei exhibiting either quasi-periodic activity or some peculiar spectral features in the ultraviolet, optical, and infrared bands, 
see e.g. \cite{Grah} and \cite{Err}. Some of the SBBH candidates can also be resolved by the present and planned mm-VLBI missions, see \cite{amalin}.

Among all SBBH candidates, the most prominent place, perhaps, 
belongs to the famous BL Lac object OJ 287 that demonstrates quasi-periodic activity with a period $\sim 11-12yr$
over the time span of a century, which may be attributed to the presence of a SBBH, \cite{Sill}. As was noted, for example, by  \cite{LV}, since the outbursts from OJ 287 often come in pairs, with individual outbursts in a pair
separated by time periods 
order of or slightly larger than a year, a natural explanation of this activity would be to associate
the outbursts with collisions between a smaller black hole (a perturber) orbiting a larger primary black hole on an eccentric orbit and an accretion disc around
the primary. In this model, the orbital plane and the disc plane are misaligned.
Since the time between the outbursts in a pair is much shorter than that  between two successive pairs, 
it is natural
to assume that the perturber-accretion disc collisions happen near the periastron of the orbit. 
Implementation of this model has led to rather extreme
parameters of the system, namely, the primary mass being as large as $20$  billion solar masses and the perturber with mass of the order of $100$ million 
solar masses having quite an eccentric orbit with eccentricity $e\approx 0.7$ and semi-major axis $a$ order of a hundred gravitational radii, see e.g \cite{V08} and \cite{V23} and references therein for a more recent discussion. Additionally, as we have mentioned above, the orbit should be significantly 
inclined with respect to the disc plane, so that its inclination angle well exceeds the disc opening angle. The model is referred hereafter to   
as PM (precessing massive) model of OJ 287.

The unusual parameters of the system naturally have led to a number of questions regarding validity of the model. For example, it was pointed out
a long time ago (e.g. \cite{IPP}) that, when a binary of unequal masses interacts with an accretion disc and 
the mass ratio, $q$, of the smaller component to the larger one is small enough, the low mass component is rather quickly dragged to the disc plane.
This happens when the perturber mass, $m$, is smaller than a typical disc mass enclosed within the orbit. On the other hand, in the opposite case  of the light disc,
it is deformed in such a way that it aligns with the orbital plane of the binary. The alignment time is again expected to be rather short. 
Both processes should operate at scales much larger than those required to explain the activity of OJ287 and, therefore, a system with the value of semi-major axis appropriate for OJ 287 is expected to have orbit and
the disc, which are aligned. Moreover, in certain cases, a cavity
in the disc inside the orbit is expected to be formed. Therefore, the existence of a system with the required properties seems unlikely from a theoretical point of view, at least, at first glance. 

Moreover, the suggested model has also been criticised from the observational
point of view. For example, it was claimed recently in \cite{Komossa} that an outburst of OJ 287 expected to occur in October 2022 according to the model did not actually happen. Based on this fact and using some empirical relationships, these authors proposed to revise the mass of the primary to a much more modest value  $\sim 10^{8}M_{\odot}$. 
In its turn, the criticism of \cite{Komossa} has prompted \cite{V23} to modify the original PM model by slightly changing the arrival times of outbursts in such a way 
that it is still able to explain the observations. 
Here we note that in order to model the accretion disc, the authors of the original PM model
used systems of non-interacting
particles (see  \cite{V23} and references therein), which may be inadequate for the description
of hydrodynamical effects in the disc.
It was also assumed that the disc is  globally flat
and resides in a plane, which coincides with the equatorial plane of the primary. 
Clearly, it is important to examine these assumptions in order to
either validate or refute this model, for example, in favour of the models
associated with jet precession (see, e.g. \cite{Br}). Of a special importance would be, for
example, to check whether or not the disc has a flat shape or it acquires a more complicated geometrical form due to secular interactions with the perturber.

{  As we discuss below, that the orbital evolution time is determined by the emission of gravitational waves in PM model alleviates the significance of the theoretical criticism. Namely,} it may then be shown that the evolution time of the bound orbit can be significantly shorter
than a typical viscous time of the accretion disc provided that the disc is geometrically thin. That means that an extended cavity in the disc cannot
be formed before the final merger of the binary and the perturber always has enough material when it intersects the disc to produce hot outflows and
the corresponding flaring activity, see e.g. \cite{IIN}. However, the timescale for alignment of the disc with the orbital plane is expected to be smaller
than the viscous time provided that the Shakura-Sunyaev parameter $\alpha$ is small, see \cite{PP83}. Therefore, the disc may be able {,   in principle,} to have enough
time to align with the orbital plane thus changing the basic assumptions of the model relying on perturber-disc collisions as the source of outbursts.
This is one of the questions we are going to address in this Paper.

In this work we apply the theory of twisted discs to find qualitative properties of the disc  shape on time scales, that
are longer than the orbital period at the  spatial scales of interest, but shorter than, or of the order of a characteristic time of orbital evolution 
due to emission of gravitational waves, $T_{GW}$. 
We assume that at the scale of interest, the disc is razor thin, with a typical ratio of the disc's half-thickness $h$ to its radius
$r$ (the disc opening angle) order of $10^{-3}$ (see e.g. \cite{SS} and \cite{IIN}) . 
We employ our model of a fully relativistic 
non-stationary twisted disc proposed in \cite{ZI} and developed 
and tested against GRMHD simulations in \cite{TFZI} and \cite{ZIFT} to see what kind of geometrical distortions could be caused by a perturber 
with the parameters of the system appropriate to PM model and how it can influence predictions of the model concerning the times of
outbursts associated with times of crossing of the disc plane by the perturber. In order to bring the problem to a treatable form and be able
to use our formalism, we make a number of assumptions. 

1) We assume that both inclinations of the orbital plane $\beta_b$ and
the disc plane $\beta(r,t)$ are formally small, which allows us to split equations describing dynamical properties of the disc into a 'background
part' and 'twisted perturbations'. For the background part, we use the \cite{NT} model and assume that the primary rotational parameter is small
enough to treat the background space-time as that of a Schwarzschild black hole. We also neglect potentially important effects associated with the disc's self-gravity. 

2) In this Paper we do not make attempts to find 
a quantitative agreement between our approach and the observations. Basic parameters of the model such as the primary and the perturber masses, orbital semi-major axis and eccentricity are 
specified according to the values proposed in \cite{V23}. Accordingly, we neglect the evolution of the orbit
due to emission of gravitational waves. The disc is initially flat, and we numerically evolve
the dynamical equations describing our model up to the end time
order of $T_{GW}$. However, we fully account for effects caused by the Einstein apsidal precession and Lense-Thirring nodal precession of the
binary orbit and of the disc rings. 

3) Since we are interested in time scales longer than the orbital time scales of the binary and of the disc, we 
double average terms in the perturbing gravitational potential, which are responsible for apsidal and nodal precession of slightly perturbed
disc rings, over the periods of the binary and of a disc ring having a given radius $r$. This is done numerically for the entire range
of $r$ under consideration. Although a similar procedure was frequently used in the case of circumbinary and circumprimary discs in protoplanetary systems (see e.g. \cite{ZL} and  \cite{Ab} for the case of eccentric circumbinary discs) we are not aware of an extensive use of this procedure \footnote{Note, however, that a similar approach was shortly discussed in \cite{HSOJZN}.} in the entire range of $r$.
 Since we use this procedure even in the region between orbital periastron and apoastron, its justification is not relied upon the usual assumption of a large mismatch between the orbital frequencies. Rather, we assume that gas elements of the disc remain on slightly perturbed circular orbits of almost the same inclination over the averaging times since the perturbing
potential is proportional to the small mass ratio $q$. 
In this way we neglect the potential influence of resonances of various kinds on the evolution of the disc rings\footnote{Our
approach is somewhat similar to the 'wire approximation' used in stellar dynamics, see e.g. \cite{RT} and \cite{GH} and references therein.}. The influence of resonances can, in principle, be done separately, see e.g. \cite{A}. 

4) In this Paper we neglect the possibility of the disc's breaking (see e.g. \cite{Y} and references therein) and the influence of non-linear effects on warp propagation (see e.g. \cite{O1}, \cite{O2} and \cite{LO}). 

5) Finally, for simplicity, we restrict our attention to the case of rather large viscosity
parameter $\alpha=0.1$, which is less numerically demanding from the computational point of view. As we are going to show, the use of a large  $\alpha$ could lead to rather non-trivial consequences.  

{  It is shown in the Paper, that the disc quickly acquires a non-trivial twisted shape, which is quasi-stationary in the frame precessing with the orbit due to the Lense-Thirring effect. Its typical inclination 
angle varies rather significantly over the scale of the order of the binary semi-major axis, with a typical
variation amplitude comparable to $\beta_b$. We provide a preliminary numerical study of the conditions of
intersections of the perturber with the twisted disc, and we show that both their number per one orbital period
and the intersection times differ significantly from the flat disc case. Thus, the effects determined by the
non-trivial disc shape should be taken into account in the modelling of the flaring activity of OJ 287 in the
framework of PM model (see
e.g. \cite{Dey} and \cite{ZM} and the references therein). They can, in principle, either refute the model
or lead to a significant modification of the model parameters\footnote{But note that a preliminary study
of the disc configurations in the case of smaller values of $\alpha$ shows that they can be quite different
from those considered in the Paper. This case is left for a future work.}.}    

The Paper is organised as follows. In Section \ref{base} we introduce basic notations, definitions, and quantities used in the following Sections as well as discuss various physical processes playing a role in our study. In Section \ref{calc} we apply our approach to the calculation of the nodal and apsidal precessions. In Section \ref{num} we present our numerical model, discuss the results of computations, and propose a simple analytical model to clarify our numerical results. Finally, in Section \ref{conc} we conclude and discuss some possible further developments of this study.

\section{Basic definitions, relations and physical processes of interest}
\label{base}

\subsection{Basic notations}\label{Basiceq}
We consider a binary black hole with two unequal masses $M$ and $m \ll M$ orbiting around each other on a slowly evolving eccentric Keplerian 
orbit with its semi-major distance being significantly larger than the gravitational radii of both components. The heavier and lighter components are referred hereafter to as the primary component and the perturber, respectively, and their mass ratio
$q={m\over M}$ is assumed to be of the order of $10^{-2}$ as suggested by the precessing binary model of OJ 287 (see e.g \cite{V23} and references
therein). It is assumed that there is a thin accretion disc around the primary component and the orbital plane is inclined with respect to
the plane of unperturbed disc.

\subsection{Characteristic temporal and spatial scales}

In our problem there are parameters and characteristic scales of two different origins, the ones associated with 
the binary and the ones associated with the disc.

\subsubsection{Characteristic scales associated with the binary}

We follow \cite{V23} 
for estimates of tentative parameters and associated characteristic scales of the binary.
Namely, we assume that the primary's mass, $M$, could be as large as $2\cdot 10^{10}M_{\odot}$, while mass of the secondary, $m$,
is about $1.5\cdot 10^{8}M_{\odot}$. This gives the mass ratio $q=7.5\cdot 10^{-3}$. We use $M_{*}=M/2\cdot 10^{10}M_{\odot}$ and
$q_{*}=q/7.5\cdot 10^{-3}$. \cite{V23} assume the orbital period in the observer frame $P_{obs}\approx 12yr$, which 
gives the orbital period in the source frame $P_{orb}\approx 9yr$ for the redshift $z=0.3$ corresponding to OJ 287. Given these
values, it is straightforward to evaluate the orbital semi-major axis, $a$. It is convenient to express it in units of the characteristic gravitational scale $r_{g}={GM\over c^2}$ and we use hereafter tilde for various spatial scales expressed in 
units of $r_g$. We have
\begin{equation}
\tilde a=60M_{*}^{-2/3}P_{*}^{2/3},  
\label{sc1}
\end{equation}
{  where $P_{*}=P_{orb}/9yr$}, and we use {  $a_{*}=\tilde a/60$} in our expressions below. 
   
Another very important quantity characterising the problem 
is orbital eccentricity, $e$. \cite{V23} argue, that the change of apsidal
angle over one orbital period, $\Delta \Psi_b$, due to Einstein precession is approximately equal to $40^{\circ}$. When 
the mass ratio $q$ is small, we can use the standard result that 
\begin{equation}
\Delta \Psi_b\approx {6\pi\over \epsilon ^2 \tilde a}, \quad \epsilon =\sqrt{1-e^2},  
\label{sc2}
\end{equation}     
to obtain {  $e=e_{*}\approx 0.7$} when $\tilde a=60$ ({  $a_{*}=1$}).

The orbital period $P_{orb}$  provides the shortest dynamical time scale at scales $r\sim a$. The corresponding mean motion, $\Omega^b_K={2\pi\over P_{orb}}\approx \sqrt{GM\over a^3}$ can be expressed as
\begin{equation}
\Omega^b_K \approx 2\cdot 10^{-8}P_{*}^{-1}s^{-1}.
\label{sc3}
\end{equation}     

The next smallest time scale is the time scale of Einstein precession of the apsidal line. It is characterised 
by a standard expression for the corresponding precessional frequency 
\begin{equation}
\Omega_{E}= \Delta \Psi_b/P_{orb}\approx 0.1\epsilon_{*}^{-2}a_{*}^{-1}\Omega^b_{K},
\label{sc4}
\end{equation}  
where $\epsilon_{*}=\epsilon/\sqrt{1-e^2_*}\approx 1.4\epsilon$. 

When the rotational parameter of the black hole, $\chi$, is not equal to zero and the orbital plane is inclined with respect
to the black hole equatorial plane, its line of nodes is
precessing with the frequency
\begin{equation}
\Omega^b_{LT}= {2\chi G^2 M^2\over c^3\epsilon^3 a^3}
\label{sc5}
\end{equation}  
due to the Lense-Thirring effect.  
Using equations (\ref{sc2}) and (\ref{sc5}) we have
\begin{equation}
\Omega^b_{LT}{  \approx }10^{-2}\chi M_{*} P_{*}^{-1}\epsilon_{*}^{-3}\Omega^b_{K},
\label{sc6}
\end{equation}
see e.g \cite{Meritt}. {  In what follows we mainly consider two absolute values of $\chi$, $|\chi|=0.5$ and $0.25$,
considering both signs of $\chi$. This sign $+$ ($-$) correspond to prograde (retrograde) black hole rotation with
respect to the motion of gas in the disc.}

Finally, the slowest orbital time scale, $T_{GW}$, is determined by emission of gravitational waves, which leads to a gradual
decrease of $a$ and $e$. We use the corresponding expression from \cite{P64} appropriate for large eccentricities
to obtain
\begin{equation}
T_{GW}\approx 10^{4}q_{*}^{-1}M_{*}^{-5/3}P_{*}^{5/3}\epsilon_{*}^{7}{\Omega^b_K}^{-1}.
\label{sc7}
\end{equation}  
In our numerical work below we consider constant values of $a$ and $e$. Thus, the evolution time of our equations describing
the disc's tilt and twist is limited from above by the expression (\ref{sc7}).

\subsubsection{Characteristic scales and times associated with the disc}

We use the standard \cite{SS} $\alpha$-model of the disc. In the framework of this model it can be shown
that the disc opening angle $\delta =h/r$, where $h$ is the disc halfthickness is nearly constant and very small, 
$\delta \approx 10^{-3}$. We use the expression for this angle provided by \cite{IIN}, hereafter IIN, which is rescaled to take
into account that typically we consider larger masses of this primary and larger accretion rate
\begin{equation}
\delta \approx  10^{-3}{\alpha_*}^{-1/10}M_{*}^{-1/10}{\dot m}^{1/5}, 
\label{sc8}
\end{equation} 
where $\alpha_*=\alpha/0.1$, $\dot m =\dot M/0.1\dot M_E$, $\dot M$ is accretion rate and $\dot M_{E}$ is its Eddington 
value defined as in INN. Background quantities of the disc, such as its surface density, accretion rate
and $\delta$ evolve on a ``viscous'' time scale 
\begin{equation}
T_{\nu}\sim \alpha^{-1}\delta^{-2}\Omega_{K}^{-1}\approx 10^{7}\alpha_{*}^{-4/5}M_{*}^{1/5}{\dot m}^{-2/5}\Omega_{K}^{-1}, 
\label{sc9}
\end{equation}  
where $\Omega_K=\sqrt{GM\over r^3}$ is the Keplerian angular frequency of a disc element on a circular orbit of radius $r$. At scales
$r\sim a$,  $T_{\nu}$ is larger than any time scale of our problem. That means that, in general, in regions where the 
perturber does not strongly
influence the disc, it is possible to consider stationary values of the background quantities provided by \cite{SS}
solution. Note, however, that this assumption breaks down in the region between the orbital periastron and  apoastron, where the perturber intersects the disc plane.

When $\alpha > \delta$ the disc tends to align with a symmetry plane of a problem on smaller scales characterised by the
corresponding characteristic alignment radius, $r_{al}$. In our case, the situation is more complicated, since there are 
two such planes - the black hole equatorial plane and the orbital plane, and, accordingly, two characteristic radii 
$r_{al,BH}$ and $r_{al,orb}$. We use the result of \cite{II} for $r_{al,BH}$,
\begin{equation}
r_{al, BH}\approx 9\cdot 10^{3}\alpha_{*}^{2/3}\chi^{2/3}\delta_{*}^{-4/3}r_{g}, 
\label{sc10}
\end{equation} 
where $\delta_{*}=\delta/10^{-3}$. For an estimate of $r_{al, orb}$ we use the result of \cite{IPP}, hereafter IPP, which 
was obtained under the assumption of a circular orbit and considering radial scales much larger than the semi-major axis. Both these
assumptions break down in our case, but, we are going, nonetheless, to use this result for a crude estimate. We have
\begin{equation}
r_{al, orb}\sim (\alpha q)^{1/2}\delta^{-1}a\approx 1.6\cdot 10^{3}(\alpha_{*}q_{*})^{1/2}\delta_{*}^{-1}a_{*}r_{g}. 
\label{sc11}
\end{equation}  
One can see that both characteristic scales can be comparable, and, therefore, one can expect relaxation of the twisted
disc to a more non-trivial quasi-stationary configuration than normally expected in the situation when either the Lense-Thirring
effect or perturbations of the disc by the binary dominate.

When $\alpha > \delta $ the disc tends to relax to a stationary configuration during characteristic time $T_{al}\sim \alpha^{-2}T_{\nu}$, see \cite{PP83}. Using (\ref {sc9}) we have
\begin{equation}
T_{al}\approx  10^{5}\alpha_{*}^{6/5}M_{*}^{1/5}{\dot m}^{-2/5}\Omega_{K}^{-1}. 
\label{sc12}
\end{equation}
Comparing (\ref{sc7}) and (\ref{sc12}) one can see that when $\alpha  \ge \sim 0.01$ $a$ and $e$ are changing 
faster than the relaxation time of the twisted disc. In this situation, on a time scale of the order of 
$T_{GW}$ effects determined by hydrodynamical interactions in the disc are expected
to be less important than the torques on the disc from the side of the binary and black hole. Since this could lead to 
some new dynamical effects this case is chosen below for our numerical and analytic studies.

\subsection{A possibility of gap formation}

Even when the orbital plane is inclined with respect to the disc plane, there are several physical processes which could lead to the formation 
of a gap in the disc in the vicinity of the orbit. 

Firstly, the perturber intersects the disc twice per orbital period at some radius $r_{int}$. Each time, the disc material inside the so-called 'accretion radius'
\begin{equation}
r_{acc}\sim 2qr_{int}
\label{rint}
\end{equation}
measured from the intersection point, is removed (see INN, IPP). This results in the outflow rate of the disc material, $\dot M_{out}$ 
order of $\sim 2\pi \Sigma r_{acc}^2/P_{orb}$, where $\Sigma$ is the disc surface density near the intersection point. Provided that
the rate of inflow, $\dot M_{in}$, of the disc material in the vicinity of the intersection point is smaller than $\dot M_{out}$ a depression 
of the surface density near $r_{int}$ is formed. As was argued in e.g. INN when $\dot M_{in}$ is associated with the accretion rate for an 
unperturbed disc even a perturber with a rather small mass can form such a depression. This effect was later observed in numerical simulations, see e.g  \cite{XP} and \cite{HSM} for the inclined case. However, in our case the orbital parameters change with time due to emission of gravitational
waves with a characteristic time, which is much smaller than the viscous time $T_{\nu}$. That means that, 
in our case in the frame comoving with 
the semi-major axis, the inflow rate can be estimated as $\sim 2\pi \Sigma r^2_{int}/T_{GW}$, where we take into account that $r_{int} \sim a$. 
In this case, from the condition $\dot M_{out} < \dot M_{in}$ we obtain
\begin{equation}
q < \sim 10^{-2}M_{*}^{5/6}P_{*}^{-5/6}\epsilon_{*}^{-7/2},
\label{rint1}
\end{equation}
where we use (\ref{sc7}).  We see that this inequality is marginally satisfied for our chosen value  $q=7.5\cdot 10^{-3}$ and, therefore,
the formation of the depression due to the mechanical withdrawal of the disc material is unlikely.

Secondly,  when a perturber corotates with the disc, an extended gap can be formed due to the presence of the Lindblad resonances. 
This effect is  expected to be very efficient for a very thin disc and quite eccentric binary, as in the situation we consider, see e.g.
\cite{AL}. Although the torque transferred to the disc by waves launching at the resonances gets smaller when orbital
inclination increases, this effect is expected to be insignificant in a situation when orbital inclination is of the order of or smaller
than orbital eccentricity, see e.g. \cite{A}. However, in our case a characteristic size of the orbit shrinks due to emission 
of gravitational waves on a time scale, which is much smaller than the viscous time $T_{\nu}$. This allows to suggest that an extended gap
is likely to be absent in our case, see e.g. \cite{AL} 
\footnote{In the case of planet migration in the protoplanetary discs, this effect was noted by \cite{HW} and later discussed analytically by e.g. \cite{LinP} and numerically by e.g \cite{MMMM} et al 2015. However, the system
under consideration differs quite significantly from those appropriate in the protoplanetary case. The issue of gap formation for the systems
of our type, which are inclined, elliptic and evolving due to a number of factors not related to the interaction with a disc, deserves a further
study.}.   

We conclude that the gas dynamical effects do not significantly change the surface density distribution. The effects related to excitation 
of waves in the disc are unlikely to produce an extended gap. On the other hand, we have checked numerically that a possible 
thin gap practically doesn't change our results. Therefore, we neglect the possibility of gap formation for our analysis below and
assume that the basic properties of the disc are given by the standard Novikov $\&$ Thorne 1973 solution.

\subsection{Coordinate systems and associated expressions}
We introduce two Cartesian coordinate systems $(X_h,Y_h,Z_h)$ and  $(X_b,Y_b,Z_b)$ centered at the position of the primary 
component and associated with the equatorial plane 
of the central black hole and the orbital plane, respectively. It is implied that the coordinates  $(X_b,Y_b,Z_b)$ are
rotated with respect to the coordinates $(X_h,Y_h,Z_h)$ by Euler angles $\beta_b, \gamma_b, \Psi_b$, where the choice of
elementary rotations is the same as in \cite{II} and \cite{DI}. {  Namely, we adopt the following convention for
a choice of rotational angles. First, $X$ and $Y$ axes are rotated by the 
angle $\gamma_b$ about the $Z$ axis. Then, we perform
rotation by the angle $\beta_b$ about the new $X$ axis, and, then rotation about the new $Z$ axis by the angle $\Psi_b$, see 
Fig. 1 of \cite{II} for a graphical representation.}

It is also implied that the periastron of the binary is situated at the $X_{b}$ axis. Thus, the angles $\beta_b$, $\gamma_b$ and
$\Psi_b$ define inclination of the orbital plane with respect to the black hole's equatorial plane, a position of the nodal line 
with respect to $X_{h}$ axis and a position of the apsidal line with respect to the nodal line. 

The position vector of the perturber has only two components: $D_x(t)$ and $D_y(t)$ in the coordinate system associated with the orbit, 
which are given by the standard Keplerian expressions
\begin{multline}
D_{x}=a(\cos E -e)\quad D_{y}=a\sqrt{1-e^2}\sin E, \\ \quad D=\sqrt{D_x^2+D_y^2}=a(1-e\cos E),  
\label{e6}
\end{multline}
where $a$ is the semi-major axis, $e$ is the eccentricity and $E $ is the eccentric anomaly, which is related to time $t$ as
\begin{equation}
t={\Omega^b_K}^{-1}(E-e\sin E).   
\label{e7}
\end{equation} 

The effects of General Relativity lead to a secular evolution of
the orbital elements. In particular, the Einstein precession and the Lense-Thirring effect lead to the evolution of $\Psi_b$ and  $\gamma_b$, respectively, with the corresponding evolution time scales $t_{E}$ and $t_{L-T}$, following from equations (\ref{sc4})
and (\ref{sc6}) above,
while emission of gravitational waves causes the evolution of $a$ and $e$, with the corresponding evolution time scale $t_{GW}$
given by equation (\ref{sc7}). 
From these equations it follows that  $t_{E}\ll t_{L-T}\ll t_{GW}$. Note that we neglect any contribution to the evolution of orbital elements due to interaction of the binary with the disc. This is justified in the case when a typical disc mass is much smaller than $m$, as
in the case of our system.  

Similarly, a disc ring can be described by the 'twisted coordinates' $(r, \Psi, \xi)$ (e.g.
\cite{Pet} and also \cite{ZI}) and references therein) with the help of two Euler angles  
$\beta (r,t)$ and $\gamma (r,t)$ that are associated with the black hole  equatorial plane. 

\subsubsection{Concrete expressions for the coordinate transformations} 

In what follows, we assume that the inclination angles $\beta$ and $\beta_b$ are small. In this situation 
one usually employs transformation laws between different coordinate systems in the linear approximation over the inclination angles,
see e.g. II. However, in our case, this leads to formally divergent expressions for apsidal and nodal precession frequencies of the disc's
rings due to the presence of the perturber, which we calculate below, see the next Section. 
This happens when $r_p \le r \le r_a$, where $r_p$ and $r_a$ are apoastron and periastron of the perturber's orbit. Since we would like to have expressions for these quantities, which are finite everywhere apart from some radius, where the perturber crosses the disc, where such divergences
are physically motivated, we have to take into account terms quadratic over the inclination angles
when considering transformations between different coordinate systems.   

More concretely, the divergences arise due to the fact, that 
in the linear approximation over  $\beta$ and $\beta_b$ 
the distance between the perturber and a particular gas element situated at the disc midplane
$\xi=0$, 
\begin{equation}
\Delta r_0\equiv \Delta r (\xi=0)=\sqrt{(D_x-X_b)^2+(D_y-Y_b)^2+Z_b^2},
\label{e1}
\end{equation},
where $X_d$, $Y_b$, $Z_b$ are coordinates of a particular element of the disc corresponding to $\xi=0$,  is formally the same as in the case when the perturber's orbital plane and the disc midplane coincide. Clearly, in this situation, $\Delta r (\xi=0)$ is equal to zero at the time 
when pertuber's distance to the origin of our coordinate systems, $D$, is the same as $r$. This happens twice per orbital
period, $P_{orb}$. In order to circumvent this obstacle, we consider the terms quadratic in $\beta$ and $\beta_b$ in the transformation laws. This results in finite distances at the times of the closest approach
 apart from a particular value of $r=r_{int}$ belonging to the interval $(r_p,r_a)$, which corresponds to the intersection of the orbital plane and the disc midplane. It is assumed below that the disc material must be removed by the action of the perturber's gravitational field close to $r_{in}$, and, therefore, our expression for the precessional frequencies remain finite at all potentially interesting radial scales.

In order to calculate  $\Delta r_0$ and other quantities interesting for us, we express 
Cartesian coordinates $(X_{b},Y_{b},Z_{b})$ in terms of the twisted coordinates 
associated with the equatorial plane,  $(r, \Psi, \xi)$ using the rotational matrices defined as in II. 
It is then convenient to represent $X_b$ and $Y_b$ in the
form  $X_b=X_0+X_1+X_2$ and $Y_b=Y_0+Y_1+Y_2$, where $X_{0}$, $Y_{0}$ do not depend on the inclination angles, $X_1$ and $Y_1$ are linear over
the inclination angles and $X_2$ and $Y_2$ are quadratic over them. Note that $X_0$, $Y_0$ and $X_2$, $Y_2$
are even with respect to $\xi$, while $X_1$ and $Y_1$ are odd. 

We easily obtain the following relations from the expressions provided in \cite{II}
\begin{equation}
X_{0}=r\cos (\tilde \gamma -\tilde \gamma_b), \quad Y_{0}=r\sin (\tilde \gamma -\tilde \gamma_b), 
\label{e3}
\end{equation} 
where
\begin{equation}
\tilde \gamma_b=\gamma_b+\Psi_b,
\quad  \tilde \gamma =\gamma +\Psi,
\label{e3a}
\end{equation} 
\begin{multline}
X_{1}=\xi(\beta \sin (\gamma -\tilde \gamma_b)+\beta_b\sin \Psi_b), \\ \quad Y_{1}=-\xi(\beta_1\cos (\gamma -\tilde \gamma_b)-\beta_b\cos \Psi_b).  
\label{e4}
\end{multline}
and
\begin{multline}
X_{2}=r(-{\beta_b^2\over 2}\sin \Psi_b \sin (\tilde \gamma -\gamma_b)+ \\ {\beta^2\over 2}\sin \Psi \sin (\gamma -\tilde \gamma_b)+\beta \beta_b \sin \Psi
\sin \Psi_b), 
\label{e3b}
\end{multline}
\begin{multline}
Y_{2}=r(-{\beta_b^2\over 2}\cos \Psi_b \sin (\tilde \gamma -\gamma_b)- \\ {\beta^2 \over 2}\sin \Psi \cos (\gamma - \tilde \gamma_b)+\beta \beta_b \sin \Psi
\cos \Psi_b). 
\label{e3c}
\end{multline}

Additionally, we have
\begin{multline}
Z_b=(1-{\beta_b^2\over 2}-{\beta_b^2\over 2}+\beta \beta_b \cos (\gamma -\gamma_b))\xi+r\beta \sin \Psi - \\ r\beta_b\sin(\tilde \gamma -\gamma_b)
+r\beta \sin \Psi.   
\label{e4a}
\end{multline}

\subsubsection{Calculation of a concrete expression for the distance between the perturber and a gas element at the disc's midplane}

Setting $\xi=0$ in (\ref{e4a}) and substituting the result and (\ref{e3}) in (\ref{e1}) we obtain
\begin{equation}
{\Delta r}^2_0=r^2+D^2-2Dr\cos(\Psi -\Psi_*)+\Delta^2_b,
\label{e10}
\end{equation} 
where 
\begin{equation}
\Delta^2_b=Dr(\beta_b\sin (\phi_{orb}+\Psi_b)-\beta \sin (\phi_{orb}+\Psi_b-\delta))^2,
\label{e10d}  
\end{equation}
$\Psi_*=\tilde \gamma_b-\gamma +\phi_{orb}$, $\delta=\gamma-\gamma_b$,
and the angle $\phi_{orb}$ is defined by the conditions ${D_x\over D}=\cos \phi_{orb}$ 
and ${D_y\over D}=\sin \phi_{orb}$. 
The correction  $\Delta^2_b$ is expected to be important only when $\Psi \approx \Psi_*$ and $r\approx D$. Therefore, we can set $D=r$ 
and $\Psi=\Psi_*$ in the expression (\ref{e10d}), and rewrite it in the form
\begin{equation} 
\Delta^2_b=r^2\beta_{eff}^2\sin^2 (\Psi_{eff}+\phi_{orb}). 
\label{e10aa}
\end{equation}
where
\begin{equation} 
\beta_{eff}=\sqrt{\beta_b^2+\beta^2-2\beta \beta_b \cos(\delta)}, 
\label{e10b}
\end{equation} 
and the angle $\Psi_{eff}$ is defined by the conditions
\begin{multline}
\cos \Psi_{eff}=(\beta_b\cos \Psi_b-\beta\cos(\Psi_b-\delta))/\beta_{eff},  \\ 
\sin \Psi_{eff}=(\beta_b\sin \Psi_b-\beta\sin(\Psi_b-\delta))/\beta_{eff}. 
\label{e10c}
\end{multline}
It follows from (\ref{e10b}) and (\ref{e10c}) that when $\beta \ll \beta_{b}$ we have $\beta_{eff}\approx \beta_b$ and $\Psi_{eff}\approx \Psi_{b}$.

Finally, we would like to take into account that disc elements with $\Delta r_{0} < r_{acc}$, where
$r_{acc}$ is given by (\ref{rint}), are removed from the disc and, accordingly, do not contribute to 
torque exerted on the disc from the side of perturbed. Taking into account again that this is important
only when $\Delta r_0$ is small, this can be done by redefinition of $\Delta^2_b$: $\Delta_b^2 \rightarrow
\Delta^2_b+r_{acc}^2$. In this way, we finally obtain
\begin{equation} 
\Delta^2_b=r^2(\beta_{eff}^2\sin^2 (\Psi_{eff}+\phi_{orb})+4q^2). 
\label{e10a}
\end{equation}

\subsection{The perturbing potential}
In general, the Newtonian gravitational potential of a binary system, with masses $M$ (primary) and $m$ (secondary) in a coordinate frame with the origin at the primary position,  has the form (see e.g. \cite{LP})
\begin{equation}
\Psi=-{GM\over |{\bf r}|}-Gm\phi_{p}, \quad \phi_{p}={1\over |{\bf r}-{\bf D}|}-{(D_xX_{b}+D_yY_{b})\over |{\bf D}|^{3}}, 
\label{e5}
\end{equation} 
where ${\bf r}$ is the radius vector of a particular gas element in the accretion disc, ${\bf D}$ {  is the vector pointing from the central black hole to the companion, with magnitude equal to the distance between the components}, and we take into account that in the coordinate frame associated with the binary its orbit lies in the plane $(X_{b},Y_{b})$.
Clearly, the last term in the expression for $\phi_p$ is determined by the non-inertial character of the chosen coordinate frame. In what follows we are going to use below
only quantities doubly averaged over the azimuthal angle $\Psi $ and the binary orbital period $P_{orb}$. One can easily see that after the averaging over the azimuthal angle the non-inertial term disappears. It is, therefore, neglected hereafter.

\section{The nodal  and apsidal evolution due to the presence of a perturber}
\label{calc}

\subsection{Calculation of nodal evolution} 
\label{nod}

In order to calculate the nodal evolution of a disc's rings due to the presence of a perturber 
we need only the $m=1$ term in Fourier decomposition of $\phi_p$ over the angle $\Psi$, which is odd with respect to coordinate $\xi$, $\phi_p^{1}$. We also average $\phi_p^1$ over binary's orbital period $P_{orb}=2\pi{\Omega^b_K}^{-1}$, thus obtaining $\bar {\phi^{1}_{p}}={1\over 2\pi a}\int dE D\phi_p^1$. We are going to denote by 
bar any other other orbital averaged quantities. Let us also remind that we are interested only in quantities, which are 
linear in the inclination angles.

Substituting (\ref{e3}-\ref{e4a}) in (\ref{e5}) we obtain
\begin{multline}
\phi_{p}=\phi_{0}+ {1\over ({\Delta r}_0)^3}(D_xX_1+D_yY_1) - \\{1\over ({\Delta r}_0)^3}(X_0X_1+Y_0Y_1+Z^2/2),
\label{e8}
\end{multline}
where $\Delta r_0$ is given by eq. (\ref{e10}), we use the correction $\propto \beta_b^2$ only in denominators in
(\ref{e8}), 
\begin{equation}
\phi_{0}={1\over \Delta r_0}
\label{e8a}
\end{equation}
determines the value of the perturbing potential in the orbital plane. It is not important for the calculation of
nodal precession, but it will be used for the calculation of apsidal precession later in Section \ref{aps}. 
Using (\ref{e3}-\ref{e4a}) it is easy to see that
$X_0X_1+Y_0Y_1+Z^2/2=0$ in the linear approximation over the inclination angles. 
Therefore, we need to evaluate and orbit average the $m=1$ term of
the expression   
\begin{equation}
\phi_{p}={(D_xX_1+D_yY_1)\over ({\Delta r}_0)^3}.
\label{e9}
\end{equation}
It is easy to see that this expression 
depends on $\Psi$ only through $\cos(\Psi-\Psi_*)$, and 
so does the $m=1$ term, $\phi_{p}^{1}$.  Therefore, for $\phi_{p}^{1}$ we obtain from (\ref{e9})
\begin{equation}
\phi^{1}_{p}={(D_{x}X_{1}+D_yY_1)\over \pi}I\cos(\Psi-\Psi_*),
\label{e11}
\end{equation}
where
\begin{multline}
I=\int^{2\pi}_{0}d\phi {\cos \phi \over (r^2+D^2-2Dr\cos \phi+\Delta^2_b)^{3/2}} = \\
{1\over (r^2+D^2)^{3/2}} J, \\ 
J =
\int^{2\pi}_{0}d\phi {\cos \phi \over (1-x\cos \phi+{\tilde \Delta_b}^2)^{3/2}},
\label{e12}
\end{multline}
where $x={2Dr\over D^2+r^2}$ and $\tilde \Delta_b^2={\Delta^2_b\over r^2+D^2}\approx{\Delta^2_b\over 2r^2}$, where we take into account
again that the correction is only important when $D\approx  r$.  
Note that (\ref{e12}) can be expressed in terms of complete elliptic integrals. Taking into account that $D_x$ is
odd and $D_y$ is even with respect to time reversal, we obtain from (\ref{e11})
\begin{multline}
\phi^{1}_{p}={(D_{x}X_{1}+D_yY_1)\over \pi}I({D_x\over D}\cos(\Psi-\gamma_{*})+{D_y\over D}\sin(\Psi-\gamma_{*}) = \\
{I\over \pi D}({D_x}^2\cos(\Psi-\gamma_{*})X_1+{D_y}^2\sin(\Psi-\gamma_{*})Y_1),
\label{e12a}
\end{multline}
where $\gamma_{*}=\tilde \gamma -\gamma=\gamma_b+\Psi_b-\gamma$, and we take into account that terms, even with respect to time reversal, disappear after averaging over the orbit to obtain the last equality.  

We substitute the explicit expressions for $X_1$ and $Y_1$ (\ref{e4}) in (\ref{e12}) and represent the result in the form
\begin{equation}
\phi^{1}_{p}={I\over \pi D}\xi({\cal A}\cos \Psi+{\cal B}\sin \Psi),
\label{e13}
\end{equation}
where
\begin{equation}
{\cal A}={1\over 2}(D^2{\cal A}_1+\Delta_D^2{\cal A}_2), \quad {\cal B}= {1\over 2}(D^2{\cal B}_1+{\Delta_D}^2{\cal B}_2),
\label{e14}
\end{equation}
${\Delta_D}^2=D_x^2-D_y^2$ and
\begin{equation}
{\cal A}_1=\beta_b\sin \delta, \quad {\cal A}_2=\beta\sin (2(\delta-\Psi_b))+\beta_b\sin(2\Psi_b-\delta),
\label{e15}
\end{equation}
\begin{equation}
{\cal B}_1=-\beta +\beta_b \cos \delta, \quad {\cal B}_2=\beta \cos (2(\delta-\Psi_b))-\beta_b\cos(2\Psi_b-\delta),
\label{e16}
\end{equation}
where $\delta=\gamma -\gamma_b$.

To perform the orbital averaging we need to integrate (\ref{e13}) over the orbit: $\bar \phi_p^1={1\over P_{orb}}\int_{0}^{P_{orb}}dt \phi^1_p={1\over 2\pi a}\int_{0}^{2\pi}dE D\phi^1_p$. From the expression (\ref{e13}) it follows 
that in order to find $\bar \phi_p^1$ we need to evaluate two quantities
\begin{equation}
K_1(r)={1\over 2\pi a}\int dE D^2I, \quad K_2(r)={1\over 2\pi a}\int dE \Delta_D^2 I.
\label{e17}
\end{equation}

Using the quantities $K_1$ and $K_2$ from the expression (\ref{e13}), it follows that orbit average perturbation of the potential
has the form
\begin{equation}
\bar \phi_1={\xi \over 2\pi }((K_1{\cal A}_1+K_2{\cal A}_2)\cos \Psi+(K_1{\cal B}_1+K_2{\cal B}_2)\sin \Psi).
\label{e17a}
\end{equation}
{  Note that it follows from the definition of $I$ (see eq. (\ref{e12})), that it has the dimension of the inverse cube of 
a distance, $I\propto r^{-3}$. Accordingly, it is seen from (\ref{e17}), that $K_1$ and $K_2$ have the dimension of the
inverse square, and, therefore, $\bar \phi_1$ had the correct dimension of the inverse distance,  $\bar \phi_1\propto r^{-1}$.}

Remembering that the perturbing gravitational potential is $-Gm\phi_{p}$, we can easily find the averaged force acting on a disc ring in the direction perpendicular to the ring plane, $F^{\xi}$, as $F^{\xi}=-Gm{d\bar \phi_1\over d\xi}$. When the influences of 
hydrodynamical interactions and the relativistic effects on motion of the ring are neglected, and the influence of the perturber is treated as perturbation, the ring remains to be circular, but the Euler angles $\beta $ and $\gamma$ are changing. It is easy
to find their evolution from the relation ${\partial U\over \partial \Psi}={1\over 2r\Omega_{K}}F^{\xi}$, where we remind that $\Omega_K=\sqrt{GM\over r^3}$, and 
$U=\dot \beta \sin \Psi -\beta \dot \gamma \cos \Psi$, see e.g. \cite{DI} and references therein. From this relation
we have
\begin{equation}
\dot \beta ={GqM\over 4\pi r\Omega_K}(K_1{\cal A}_1+K_2{\cal A}_2),\quad \beta \dot \gamma ={GqM\over 4\pi r\Omega_K}(K_1{\cal B}_1+K_2{\cal B}_2),
\label{e17b}
\end{equation}

\begin{figure}
\begin{center}
\includegraphics[width=8.0cm,
angle=0]{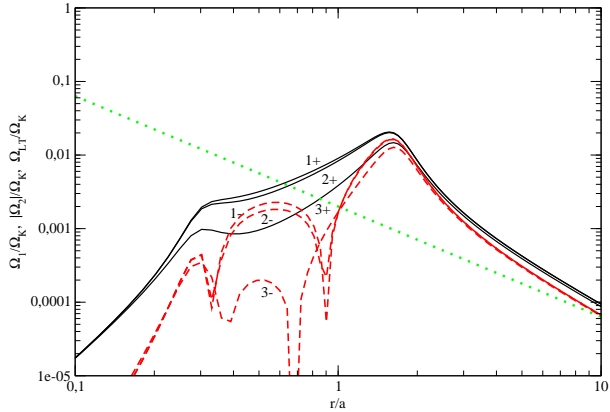}
\vspace{1cm}
\caption{Solid, and dashed curves represent $\Omega_1$ and $\Omega_2$ in units of $\Omega_K$ as functions of $r$, while the dotted curve
is the Lense-Thirring frequency (in units of $\Omega_K$) of a disc element on a circular orbit. Since $\Omega_2$ changes its sign, only
the absolute value is shown. Note that it is positive in both asymptotic limits. Solid and dashed curves with larger values of the argument
correspond to smaller $\beta_b$. {  We show $\beta_b=0.01$ labeled as $1\pm$ just above the curves with $\pm$ corresponding to
$\Omega_{1,2}$, respectively. $\beta_b=0.1$ and $0.5$ are labeled as $2\pm$ and $3\pm$ just below the curves with the same sign
convention as in the previous case.} $\Psi_{eff}=0$ for all curves.}
\label{fig2}
\end{center}
\end{figure}

It is useful to calculate the corresponding time derivative of a complex quantity
${\bf W}=\beta e^{i\gamma}$. Using eqns. (\ref{e15}), (\ref{e16}) and (\ref{e17b}) we easily
obtain
\begin{equation}
\dot {\bf W}\equiv \dot {\bf W}_b=i(\Omega_1({\bf W}_b-{\bf W})+\Omega_2e^{2i\Psi_b}(e^{2i\gamma_b}{\bf W}^{*}-{\bf W}_b)), 
\label{e17c}
\end{equation}
where
\begin{equation}
\Omega_{1,2} ={GqM\over 4\pi r\Omega_K}K_{1,2},
\label{e17d}
\end{equation}
${\bf W}_b=\beta_b e^{i\gamma_b}$ and $*$ stands for the complex conjugate. We
show $\Omega_1$ and the absolute value of $\Omega_2$ together with Lense-Thirring frequency of a circular disc element\footnote{This is obtained from (\ref{sc6}) by setting $r=a$, $\Omega_K=\Omega^b_K$ and using $\epsilon_{*}=1.4$, which corresponds the circular value of 
$\epsilon=1$.} for $e=0.7$ and $\beta_b=0.01$, $0.1$ and $0.5$ in Fig. \ref{fig2}. It is seen that the curves corresponding to $\beta_b=0.01$
and $0.1$ are close to each other. This is because when $\beta_b$ is small, a contribution of the term proportional to $\beta_b$ to
$\Delta_b$ in eq. (\ref{e10a}) plays a smaller role than the term determined by the accretion radius, which depends only on the mass
ratio.
\bigskip 
\bigskip
{  It is seen from eq. (\ref{e17c}) that the term proportional to $\Omega_1$ causes precession of a disc ring
about the axis perpendicular to the orbital plane. In principal, when the second term proportional to $\Omega_2$ 
is also taken into account, the ring dynamics could be non-trivial. As was discussed in e.g. \cite{Ab}, \cite{Aly}
and \cite{ZL} when precession of $\Psi_b$ is solely determined by the binary, under certain circumstances, it
is possible to obtain configurations of free particles and particles in a disc precessing about the direction 
pointing towards the periastron of the orbit. However, in our case, the Einstein precession of $\Psi_b$ appears to
be much faster than the one determined by the binary. The term proportional to $\Omega_2$ in this situation 
is averaged out and causes little effect on the evolution of our system, see below.}

\subsubsection {Expressions for $K_1$ and $K_2$ in asymptotic limits  $r/a \ll 1$ and $r/a \gg 1$} 
\label{limits}

In both limits $r/a \ll 1$ and $r/a \gg 1$, it is easy to evaluate $K_1$ and $K_2$ analytically. For that, we note that in these limits
we can set $\Delta_b^2=0$ and that the integral
$J$ defined in (\ref{e12}) is approximately equal to ${3\pi x\over 2}$ in both limits. 
Using these facts, it can be shown that integrals entering
(\ref{e17}) are elementary when $r \gg a$ and they can be shown to be reduced to Legendre polynomials of $\epsilon =\sqrt{1-e^2}$ when $r \ll a$. In the first case, we obtain
\begin{equation}
K_1(r)\approx {3\pi a^2\over r^4}(1+{3\over 2}e^2), \quad K_2(r)\approx {15\pi \over 2}e^2{a^{2}\over r^4},
\label{e18}
\end{equation}
while in the second case, we have
\begin{equation}
K_1(r)\approx {3\pi r\over a^3\epsilon^3}, \quad K_2(r)\approx 0.
\label{e19}
\end{equation}
Substituting these expressions in (\ref{e17a}) we obtain
\begin{multline}
\bar \phi_{p}^{1}\approx {3\over 2}\xi {a^2\over r^4}(1+{3\over 2}e^2)({\cal A}_1\cos \Psi+{\cal B}_1\sin \Psi)+ 
\\
{15\over 4}e^2\xi {a^2\over r^4}({\cal A}_2\cos \Psi+{\cal B}_2\sin \Psi)
\label{e20}
\end{multline}
when $r \gg a$ and 
\begin{equation}
\bar \phi_{p}^{1}\approx {3\over 2}\xi {r\over a^3\epsilon^3}({\cal A}_1\cos \Psi+{\cal B}_1\sin \Psi)
\label{e21}
\end{equation}
when $r \ll a$.  

Substituting (\ref{e18}) into (\ref{e17b}) and using (\ref{e15}) and (\ref{e16}), it is straightforward to show that equations 
(\ref{e17b}) are equivalent to equations (3) of \cite{ZL} when $\beta$ and $\beta_b$ are small.

\subsection{The rate of apsidal precession}
\label{aps}

In order to calculate the apsidal advance rate, $\nu$, at first we calculate the angular frequency of circular motion, perturbed
by the presence of the companion, $\Omega$,
according to the relation 
\begin{equation}
\Omega^2={GM\over r^3}-{Gm\over r}{d\phi_p\over dr},
\label{e25}
\end{equation}
and epicyclic frequency, $\kappa$, which follows from the relation 
\begin{equation}
\kappa={2\Omega\over r}{d\over dr}(r^2\Omega).
\label{e26}
\end{equation}
Then, the apsidal advance rate is determined by their difference, $\nu=\Omega-\kappa$. Assuming that the perturbing potential
is small in absolute value, we can easily obtain
\begin{equation}
\nu={Gm\over 2\Omega_K}({d^2\bar \phi_p\over dr^2}+{2\over r}{d\bar \phi_p\over dr}),
\label{e27}
\end{equation} 
where 
the bar stands again for the double average 
over the azimuthal angle and the orbit. 

For our purposes, it is sufficient to use in (\ref{e27}) the doubly averaged value of the perturbing potential in the orbital plane
(\ref{e8a}). Thus, we neglect any contribution to the apsidal advance rate due to the difference between orbital plane and
the plane of the disc. This value can be represented in the form
\begin{equation}
\bar \phi_p ={1\over 2\pi}\int_0^{2\pi}dE {(1-e\cos E)\over \sqrt{r^2+D^2}}J(x),
\label{e28}
\end{equation} 
where $J(x)$ is given in (\ref{e12}).

Note that we assume that the disc material always moves in the positive direction, while the motion of the perturber can
be both in positive and negative directions. In the latter case, (\ref{e27}) should be multiplied by minus.

\begin{figure}
\begin{center}
\includegraphics[width=8.0cm,
angle=0]{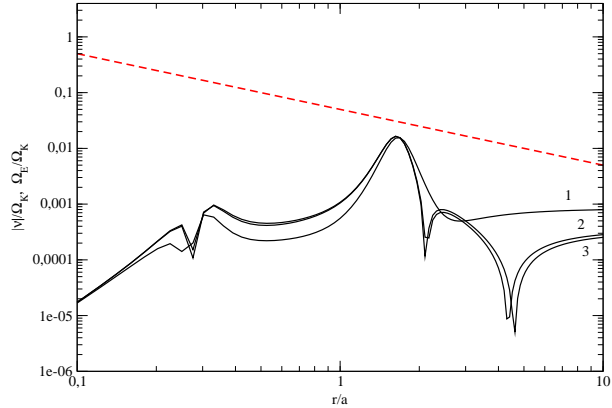}
\vspace{1cm}
\caption{The absolute value of $\nu$ in units of $\Omega_K$ is shown as a function $r/a$ together with the Einstein precession frequency
in the same units represented by the dashed curve. Curves with larger values of the argument correspond to smaller values of $\beta_b$. Note 
that at large values of $r/a$ $\nu$ is negative. {  We label the curves corresponding to $\beta_b=0.01$ and $0.1$ by $1$ and $2$ just above the
curves, and the curve corresponding to $\beta_b=0.5$ is labeled as $3$ just below the curve.}}
\label{fig3}
\end{center}
\end{figure}

We show the absolute value of $\nu$ in Fig. \ref{fig3} for $e=0.7$, $\Psi_{eff}=0$ and $\beta_b=0.01$, $0.1$ and $0.5$ 
together with the Einstein apsidal precession frequency. It is seen that Einstein's precession dominates at radial scales of interest. Therefore,
for systems with the considered parameters, the contribution of gravitational potential of the perturber is rather insignificant.
\bigskip   
\bigskip

\section{Numerical and analytic analysis of the behaviour of the twisted disc in the case of relatively large viscosity}
\label{num}
As we have mentioned above, in this Paper we consider only the discs with a relatively large viscosity, $\alpha \sim 0.1$.
This is because they are expected to evolve in manner, which is qualitatively different from the low viscosity case, since
their alignment time $T_{al}$ is smaller than the system life time $T_{GW}$. We also fix the orbital parameters of the binary
to the values introduced above and consider two values of the binary's inclination $\beta_b=0.1$ and $0.5$.

\subsection{An analytic model of the disc evolution}
\label{analyt}
\subsubsection{A simple qualitative analytic model}
\label{analyt1}

Since we are going to consider the disc at times less than $T_{al}$, in a zero approximation, it appears to be reasonable to neglect
hydrodynamical interactions in the disc at all and assume that the disc is evolving only due to the torques exerted by the binary
and black hole   
\begin{equation}
\dot {\bf W} =i\Omega_{LT}{\bf W}+\dot {\bf W}_b,
\label{e29}
\end{equation} 
where we remind that ${\bf W}=\beta(t,r)e^{\gamma (t,r)}$, 
$\Omega_{LT}={2\chi  G^2 M^2\over c^3r^3}$ is the Lense-Thirring frequency and $\dot {\bf W}_b$ is given by eq. (\ref{e17c}). 
Since $\dot {\bf W}_b$ depends on $\gamma_b$ and $\Psi_b$ as well as on $\beta$ and $\gamma$ through the dependencies
of the frequencies $\Omega_1$ and $\Omega_2$ on the disc's Euler angles, eq. (\ref{e29}) is a rather complicated non-linear
equation with time dependent coefficients. To make it analytically treatable, at first we average $\Omega_1$ and $\Omega_2$ 
over the angle $\Psi_{eff}$, thus considering $\bar \Omega_{1,2}={1\over 2\pi}\int^{2\pi}_{0} d\Psi_{eff} \Omega_{1,2}$ and
set $\beta=\beta_b$ in the resulting expressions. Secondly, we neglect the term proportional to $\Omega_{2}$ in (\ref{e17c}).
After making these simplifications, eq. (\ref{e29}) becomes elementary
\begin{equation}
\dot {\bf W} =i(\bar \Omega_1\dot {\bf W}_b+(\Omega_{LT}-\bar \Omega_1){\bf W}),
\label{e30}
\end{equation}  
with the solution
\begin{multline}
{\bf W}={\bar \Omega_1\beta_b \over \Delta \Omega }e^{i\Omega^b_{LT}t}(1-e^{i(\Omega_{LT}-\bar \Omega_1-\Omega^b_{LT})t}), \\ 
\Delta \Omega =\Omega^b_{LT}-\Omega_{LT}+\bar \Omega_1,
\label{e31}
\end{multline}
subject to the initial condition ${\bf W}(r,t=0)=0$,
where $\Omega^{b}_{LT}$ is given by (\ref{sc5}). Equation (\ref{e30}) has a resonance at $r=r_r$, such as $\Delta \Omega (r_r)=0$. At radii $r\sim r_r$, 
the assumptions leading to (\ref{e30}) are broken and a more accurate treatment is required, {  see below}. 

When $|\Delta \Omega |t\gg 1$ the last term in  the brackets in (\ref{e31}) has a large phase. Since $\Delta \Omega $ is a function of $r$, this leads to very sharp
variations of this term over $r$ at a given time, which are supposed to be smeared out by the action of
hydrodynamical interactions. We average over these variations, thus neglecting this term. In the end, we have 
an expression for a twisted disc, which is precessing with the Lense-Thirring frequency $\Omega^{b}_{LT}$ and is approximately stationary
in the precessing frame, with the angle $\beta$ varying with $r$ as
\begin{equation}
\beta ={\bar \Omega_1\beta_b \over \Delta \Omega }.
\label{e32}
\end{equation}
We compare below the expression (\ref{e32}) with the results of numerical simulations.

\subsubsection{A more accurate approach to the calculation of the quasi-stationary shape of the disc in the precessing 
frame}
\label{analyt2}
Now let us consider the disc shape near $r_r$. It follows from eq. (\ref{e32}) that $\beta$ formally diverges at this radius.
In order to remove this divergency, we take into account hydrodynamical interactions between the disc's rings. For that we use
an appropriate modification of the simplest form of dynamical equations describing twisted discs proposed in \cite{DI}, see their equations (38) and (39). Note that these equations are approximately valid in the limits $r \gg 1$ and $\alpha \ll 1$. They contain  
the variable ${\bf W}$ and an auxiliary complex variable ${\bf A}$. Assuming that the disc is stationary in the frame precessing with Lense-Thirring frequency of the binary, $\Omega^{b}_{LT}$, these variables can be represented as ${\bf W}=\hat {\bf W}(r)e^{i\Omega^b_{LT}}$ and ${\bf A}=\hat {\bf A}(r)e^{i\Omega^b_{LT}}$.
Instead of the last term in eq. (39) of \cite{DI} we use eq. (\ref{e30}).
In eq. (38) of \cite{DI} we neglect the last term in the bracket on r.h.s, express $\hat {\bf A}$ in terms of $\hat {\bf W}$
and substitute the result in eq. (39). The resulting equation can be further simplified in a region of $|r-r_r| \ll r_r$. We represent $\Delta \Omega$ in this region as
\begin{equation}
\Delta \Omega =-\Omega_{*}x, \quad x={r-r_r\over r_r},
\label{w1}
\end{equation}
where $\Omega_{*}=-r_r{d\Delta \Omega \over dr}(r=r_r)$, 
and introduce two 'renormalised' dimensionless frequencies $\omega_{*}$ and $\omega_1$, according to the rule
\begin{equation}
\omega_{*}={4\alpha \over \delta^2}{\Omega_{*}\over \Omega_{K}(r_r)}, \quad \omega_{1}={4\alpha \over \delta^2}{\bar \Omega_{1}(r_r)\over \Omega_{K}(r_r)}.
\label{w2}
\end{equation}
It is evident from their definitions that both $\omega_{*}$ and $\omega_{1}$ do not depend on $r$.
  
In this way, we obtain
\begin{equation}
{d^2\over dx^2}\hat {\bf W}+i(\omega_1\beta_b-\omega_{*}x\hat {\bf W})=0.
\label{w3}
\end{equation}  
It is easy to see from (\ref{w3}) that it can be reduced to a standard form of inhomogeneous Airy equation by rescaling
$\hat {\bf W}$ and $x$. Since $\omega_*$ and $\omega_1$ are proportional to $\delta^{-2} \sim 10^{6}$ they are expected 
to be quite large. That means that when $|x| \gg x_{*}$, where $x_{*}\ll 1$ is specified below, 
the first term on r.h.s. of (\ref{w3}) can be neglected and we have
\begin{equation}
\hat {\bf W}={\omega_1\over \omega_{*}x}\beta_b.
\label{w4}
\end{equation}  
Using (\ref{w1}) and (\ref{w2}) it is evident that (\ref{e32}) tends to (\ref{w4}) in the limit $x\rightarrow 0$. Thus,
we are looking for a solution to (\ref{w3}), which must tend to (\ref{w4}) in the case $|x|\gg x_{*}$.

One can see that such a solution can be expressed in terms of Scorer's function $Hi(z)$ (see e.g. \cite{Olv} for its
definition and properties) of a complex argument
\begin{equation}
\hat {\bf W}=i\pi{\omega_1\over \omega^{2/3}_{*}}\beta_b Hi (-i\omega_{*}^{1/3}x).
\label{w5}
\end{equation}
Note that $Hi(z)$ admits the integral representation (e.g. \cite{Olv})
\begin{equation}
Hi (z)= {1\over \pi}\int_{0}^{\infty}dt \exp (-{t^3\over 3}+zt),
\label{w6}
\end{equation} 
which allows for a very efficient numerical evaluation of (\ref{w5}). The scale $x_*$ can be defined by the condition that
the absolute value of the argument in (\ref{w5}) is equal to one. From that we have $x_{*}=\omega_{*}^{-1/3}$. 

Now we change the dependent variable $x$ in (\ref{w5}) to $\tilde x=-{\omega_1\over \omega_*}{\Delta \Omega (r)\over \bar  \Omega_1(r)}$,
where $\Delta \Omega (r)$ and $\bar \Omega_1 (r)$ are understood to be functions of an arbitrary $r$. 
When $r \rightarrow r_r$ $\tilde x \rightarrow x$ by its construction. The expression
\begin{equation}
\hat {\bf W}=i\pi{\omega_1\over \omega^{2/3}_{*}}\beta_b Hi (-i\omega_{*}^{1/3}\tilde x)
\label{w7}
\end{equation} 
is reduced to (\ref{w5}) when $|x|\ll 1$ and it is reduced to (\ref{e32}) when $|x| \gg x_{*}$. It provides an approximate
analytic solution to our problem valid for any value of radius.  We compare our analytic expression (\ref{w7}) with the results
of our numerical work below, see Section \ref{comp}.

\subsection{Numerical calculations}

\subsubsection{Specific details of our numerical approach}

The hydrodynamical model of a twisted disc which includes the gravitomagnetic action of 
the primary black hole is considered here employing the dynamical equations derived in 
\citet{ZI}. The equations (60-61) of \citet{ZI} describe the geometrically thin
relativistic accretion disc around a slowly rotating black hole. These equations are linear 
with respect to the disc inclination angle and based on the background model of the
vertically isothermal flat disc as introduced by the Novikov-Thorne solution. 
It is assumed that the disc aspect ratio is determined by the Thomson scattering, 
which yields its profile according to equation (37) of \citet{ZI}. 
In the results shown below, it is assumed that the characteristic aspect ratio 
entering equation (37) of \citet{ZI}, $\delta_*=0.001$.
The standard form of the viscosity prescription is given by (35) of \citet{ZI}, which
introduces the dimensionless viscosity parameter $\alpha$. 

In this Paper, 
we incorporate the new terms entering the right-hand side of equation 
(\ref{e17c}) into the right-hand side of the twist equation (61) of \citet{ZI}. 
Also, the correction to epicyclic frequency due to the gravitational 
influence of the secondary black hole, see equation (\ref{e27}), 
is taken into account in equation (60) of \citet{ZI}, which describes the velocity 
perturbations induced by the disc twist.

The dynamical equations generalised in this way are numerically advanced 
using an explicit second-order scheme. The dynamical variables ${\bf W}$ and 
${\bf B}$\footnote{These notations are taken from \citet{ZI}, the variable ${\bf B}$ is related to
the variable ${\bf A}$ introduced in \cite{DI} and
discussed in Section \ref{analyt2}. In particular, we have ${\bf B}\approx-2i{\bf A}$ far from the black hole.} are set on the spatial grids
uniform in $\sqrt{r}$ with the grid nodes for ${\bf B}$ shifted with respect to the grid nodes for ${\bf W}$ by a half step. 
The quantities $K_1$, $K_2$ and $\nu$ representing the influence of the pertuber are tabulated first on the separate three-dimensional mesh of its arguments, which are 
$\beta_{eff}$, $\psi_{eff}$ and $r$.
Next, they are evaluated in the grid nodes of the scheme by the successive linear interpolations along $r$, $\psi_{eff}$ and $\beta_{eff}$. Since the disc shape 
and the binary orbit evolve, this procedure is repeated at each slice.
The time step is limited by a value inverse to the largest absolute value of the 
frequency being the solution to local dispersion relation derived from the dynamical 
equations. We performed additional tests of the scheme in order to check that the convergence
of the numerical solution is achieved for the case $\alpha=0.1$ and, additionally, for 
$\alpha=0.01$. For that, the spatial grid step must be significantly
smaller than the disc thickness in the vicinity of the inner boundary in the case $\alpha=0.01$. However, it may even exceed the inner disc thickness in the case $\alpha=0.1$.
Thus, we find that the case of smaller viscosity is far more demanding for the scheme.
The regularity boundary conditions, ${\bf B}=0$ and 
$d{\bf W}/{dr}=0$, are imposed close to the inner boundary of the disc, where the surface 
density of the standard accretion disc vanishes. 
Note that in the limit of slow primary black hole
rotation, the inner boundary of a twisted disc is formally set to its Schwarzschild value, 
$r=6 GM/c^2$, see the analysis in \citet{ZI}. 

\subsubsection{Results of numerical calculations of the case with $\alpha=0.1$}

We consider numerical runs with $\chi=\pm 0.5, \pm 0.25$, and 
$\beta_b=0.1, 0.5$. 
We evolve all runs up to the end times order of 
or larger than the evolution time $T_{GW}$.
\bigskip
\begin{figure}
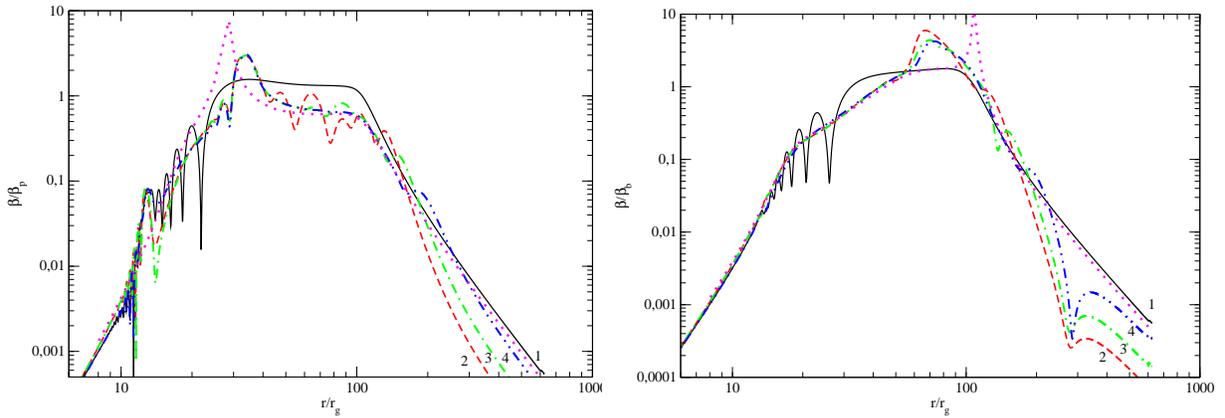

\begin{center}
\includegraphics[width=8.0cm,
angle=0]{n05bet01pl.eps}
\hspace{0cm}\includegraphics[width=8cm,
angle=0]{n05bet01mn.eps}
\vspace{1cm}
\caption{Left panel. The numerical dependencies of $\beta/\beta_b$ on $r$ for different times as well as the analytic dependency (\ref{e32}), see the text
for a description of particular curves. The case $\beta_b=0.1$ and $\chi=0.5$ is shown. Right panel. The same as the left panel, but for the retrograde
black hole rotation $\chi=-0.5$.}
\label{fig4}
\end{center}
\end{figure}
\bigskip

\begin{figure}
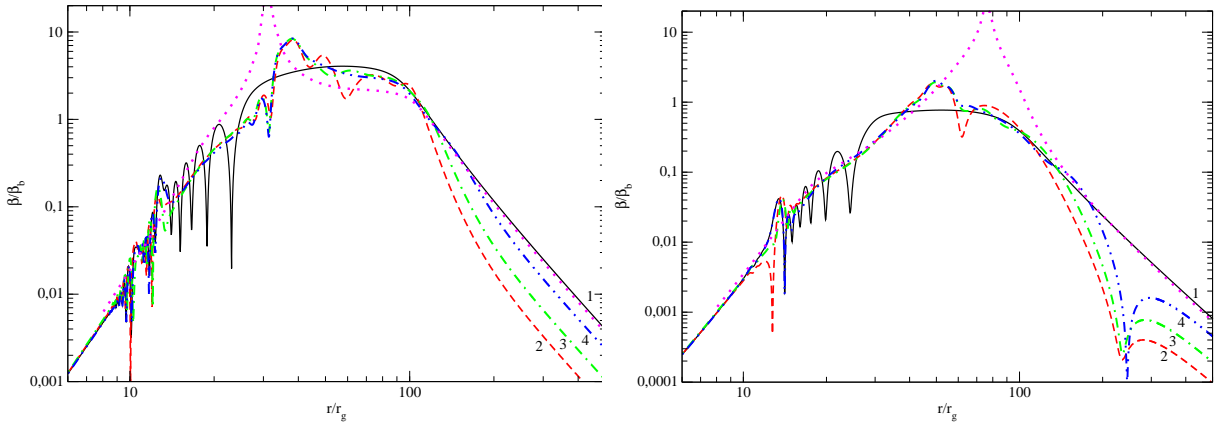

\begin{center}
\includegraphics[width=8.0cm,
angle=0]{n05bet05pl.eps}
\hspace{0cm}\includegraphics[width=8cm,
angle=0]{n05bet05mn.eps}
\vspace{1cm}
\caption{The same as Fig. \ref{fig4}, but for the case $\beta_b=0.5$.}
\label{fig5}
\vspace{1cm}
\end{center}
\end{figure}
\bigskip

In Figs.\ref{fig4} and \ref{fig5} we show ratios $\beta/\beta_p$ on $r$ for different times $t$, for the case $|\chi=0.5|$, for
$\beta_b=0.1$ and $\beta_b=0.5$, respectively. {  Solid (labeled as 1), dashed (labeled as 2), dot dashed (labeled as 3) and dot dot dashed (labeled as 4) curves correspond to $t=215{\Omega_K^b}^{-1}$,
 $t=2155{\Omega_K^b}^{-1}$,  $t=4310{\Omega_K^b}^{-1}$ and  $t=10775{\Omega_K^b}^{-1}$, respectively. Note that all labels are always
in the bottom right corner, the label 1 is always slightly above the corresponding curve, while all other labels are slightly below.}

One can see from these Figs. that 
the distribution of the disc inclination angle relaxes in a relatively short time to a quasi-stationary shape. The simple analytic expression (\ref{e32}) {  shown by the dotted curve} describes rather well this distribution apart from the region, where $\beta$ is close to its maximal value and $r\sim r_r$ {  as expected}.
The agreement is better for the prograde black hole rotation and smaller $\beta_b$.
\bigskip
\bigskip
\begin{figure}
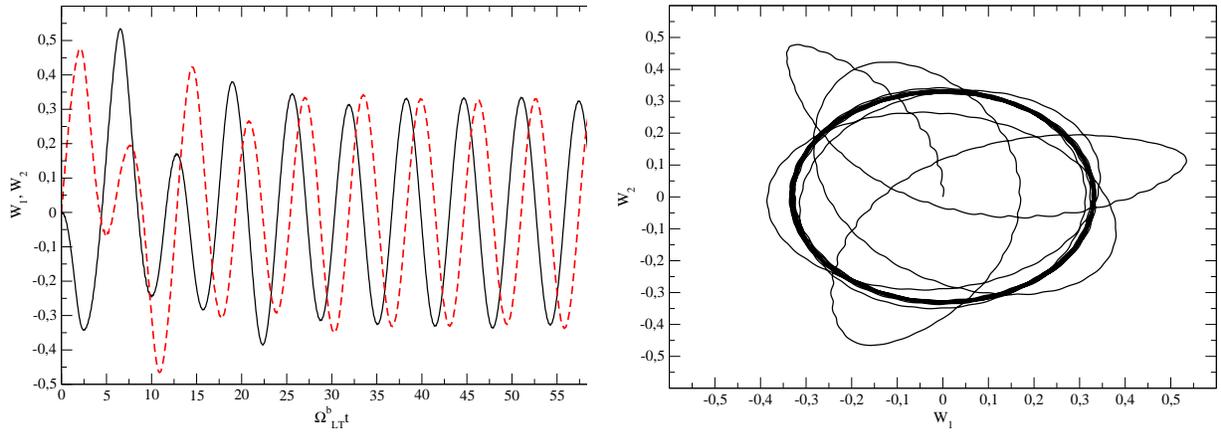

\begin{center}
\includegraphics[width=8.0cm,
angle=0]{nWt.eps}
\hspace{0cm}\includegraphics[width=8cm,
angle=0]{nW1W2.eps}
\vspace{1cm}
\caption{Left panel. $W_1(r=a)$ and $W_2(r=a)$ as functions of time in units ${\Omega_{LT}^b}^{-1}$. The case $\chi=0.5$ is shown. 
Right panel. The corresponding evolutionary track on the plane $(W_1,W_2)$. }
\label{fig6}
\end{center}
\end{figure}
\bigskip
The dependency of $W_1={\it Re }({\bf W})=\beta \cos \gamma $ and $W_2={\it Im }({\bf W})=\beta \sin \gamma $ at the radius $r=a$ 
on time in units of the inverse Lense-Thirring frequency of the binary ${\Omega_{LT}^b}^{-1}$ is shown in the left panel of Fig. \ref{fig6}, while the
corresponding evolutionary track on the plane $(W_1,W_2)$ is shown on the right panel of the same Fig. One can see that at sufficiently large
values of  ${\Omega_{LT}^b}t$ the evolution is practically harmonic, with the phase of $W_1$ differing from that of $W_2$ by $\pi/2$ and the time period
approximately equal to $2\pi {\Omega_{LT}^b}^{-1}$. We have checked that the same type of evolution is observed for other radii and other calculation parameters.
This confirms that the disc is indeed precessing with its precessional frequency equal to the Lense-Thirring frequency of the binary as suggested by our analytic
model.  

\bigskip
\begin{figure}
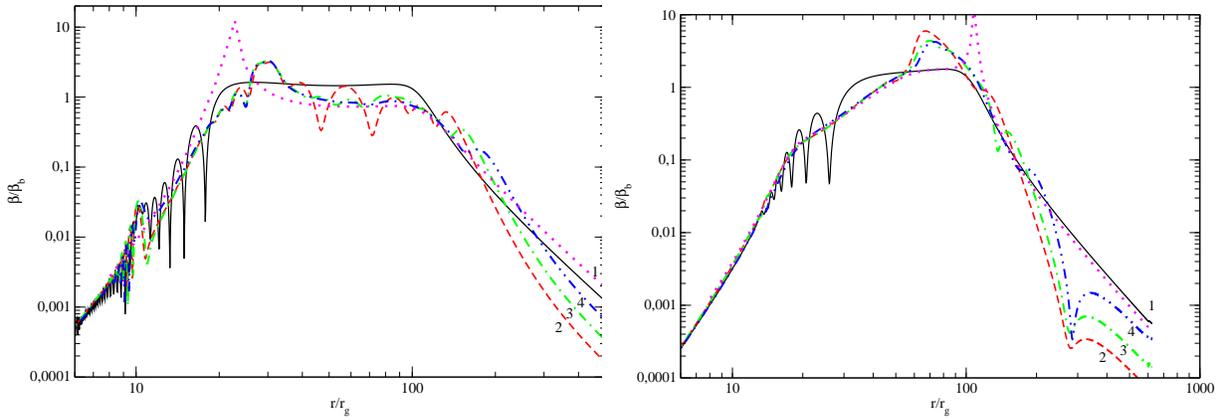

\begin{center}
\includegraphics[width=8.0cm,
angle=0]{n025bet01pl.eps}
\hspace{0cm}\includegraphics[width=8cm,
angle=0]{n05bet01mn.eps}
\vspace{1cm}
\caption{The same as Fig. \ref{fig4}, but for smaller $|\chi|=0.25$. Also note that the dot dot dashed curves correspond to a slightly smaller  $t=8620{\Omega_K^b}^{-1}$ in comparison to the analogous curves in Fig. \ref{fig4}. }
\vspace{1cm}
\label{fig7}
\end{center}
\end{figure}
\bigskip
\bigskip
\begin{figure}
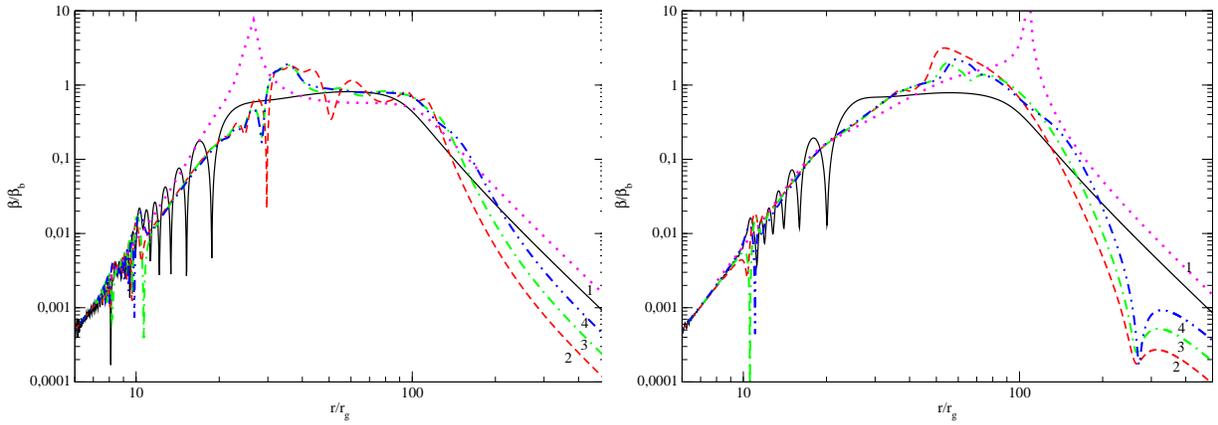

\begin{center}
\includegraphics[width=8.0cm,
angle=0]{n025bet05pl.eps}
\hspace{0cm}\includegraphics[width=8cm,
angle=0]{n025bet05mn.eps}
\vspace{1cm}
\caption{The same as Fig. \ref{fig7}, but for the case $\beta_b=0.5$.}
\vspace{1cm}
\label{fig8}
\end{center}
\end{figure}
\bigskip
The case $|\chi|=0.25$ is qualitatively similar to the case $|\chi|=0.5$. Therefore, we show only the dependencies of $\beta$ on $r$ for particular moments of
time together with the analytical curves {  represented again by dotted lines}. These are shown in Figs. \ref{fig7} and \ref{fig8}, which are analogous to Figs. \ref{fig4} and \ref{fig5}.
As seen from these figures, the agreement between analytical and numerical results looks slightly better for the smaller absolute value of $\chi$. 

\subsubsection{A more detailed comparison of our analytic and numerical results}
\label{comp}  
\begin{figure}
\begin{center}
\vspace{0cm}
\includegraphics[width=8.0cm,height= 9.0cm,angle=0]{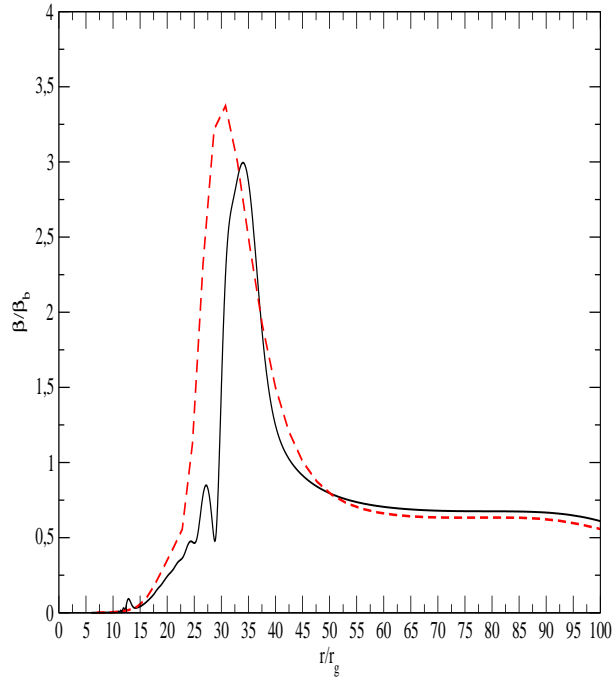}
\vspace{1cm}
\caption{The ratio of the inclination angle $\beta$ to the binary inclination angle $\beta_b$ is shown as a function
of $r$. The solid and dashed curves represent the result of a numerical calculation and our analytic expression (\ref{w7}),
respectively. Note we use $\chi=0.5$, $r_r\approx 0.5a$, $\Omega_{1}(r_r)\approx 2.9\cdot 10^{-3}\Omega_{K}^{b}$ and
$\Omega_{*}\approx 1.4\cdot 10^{-2}\Omega_{K}^b$ to plot the analytic curve.}
\vspace{1cm}
\label{fig9}
\end{center}
\end{figure}

Since the expression for $\Omega_1$ given by equation (\ref{e17d}) depends both on $r$ and the angle $\Psi_{eff}$, in order to
compare our results with the analytic expression (\ref{w7}), we remind that we average numerically calculated profiles of $\Omega_{1}$ over this angle in order to obtain $\bar \Omega_{1}$ entering (\ref{w7}). 
We use the case $\beta_b=0.1$ and $\chi=0.5$ for the comparison, the result is shown in Fig. \ref{fig9} for the case $\beta_b=0.1$ together with the corresponding result of numerical computation taken at some late computational time when the numerical distribution of $\beta$ is almost stationary. As seen from this figure, the numerical and analytic curves almost coincide at radii larger than the radii corresponding 
to the maximum of $\beta$. These maximal values agree with the accuracy order of 10 per cent. Also, the radius corresponding to the
maximum value of the analytic curve is shifted towards smaller values of $r/r_{g}$ by the same amount. Overall, we see that our rather
simple analytic approach gives quite a reasonable approximation to the numerical result.
\bigskip    
\bigskip

\subsection{The influence of the disc tilt and twist on timing of intersections between the perturber and the disc}
\label{inter}


Now let us briefly consider how possible conditions of the intersections between the perturber and the disc are modified
by the presence of the disc's tilt and twist. For that, we look for the intersections of the disc by the pertuber by comparing 
the vertical coordinate of the perturber in the coordinate system associated with the primary, $Z_h^{b}$, calculated along
its orbit, with the corresponding height of the disc,  $Z_h^{d}$. 
$Z_h^{d}(t)$ is strictly defined as the vertical coordinate of a point on 
the disc belonging to the line perpendicular to the equatorial plane, which crosses the orbit at the point with the coordinate 
$Z_h^{b}(t)$. Since both quantities are taken along the orbit, it is clear that the perturber crosses the disc when and only when $Z_h^b(t)=Z_h^d(t)$. It is also evident that when the disc is flat the condition
of intersection is just $Z_h^b(t)=0$.

The instant location of the perturber with respect to the ascending node of its orbit 
is evaluated as $\Psi = \Psi_b + n_b$, where $n_b$ is the current true anomaly of the perturber.
The latter is determined numerically employing an iterative solution 
of the Keplerian equation (\ref{e7}) and the standard relation between the true and the 
eccentric anomalies. We assume below that $n_b(t=0)=0$, and that $n_b=0$ corresponds 
to the passage of periastron.
In the linear approximation over the angle $\beta_b$, we have 
\begin{equation}
\label{Z_b}
Z_h^b = r_b \beta_b \sin \Psi, 
\end{equation}
where $r_b(n_b)$ is the current distance between the holes.

Similarly, in the linear approximation over $\beta$, we have
\begin{equation}
\label{Z_d}
Z_h^d = r_b \beta(r_b) \sin \Psi_{tw}, 
\end{equation}
where $\Psi_{tw}$ is the position angle of the disc element located above (or below) the perturber
with respect to the ascending node of the disc ring specified by the angles 
$\beta(r_b)$ and $\gamma(r_b)$. In the same approximation we can use
\begin{equation}
\label{intsn_ang}
\gamma(r_b) + \Psi_{tw} = \gamma_b + \Psi,
\end{equation}
which determines $Z_h^d$ together with equation (\ref{Z_d}).

\begin{figure}
\begin{center}
\includegraphics[width=16.0cm,
angle=0]{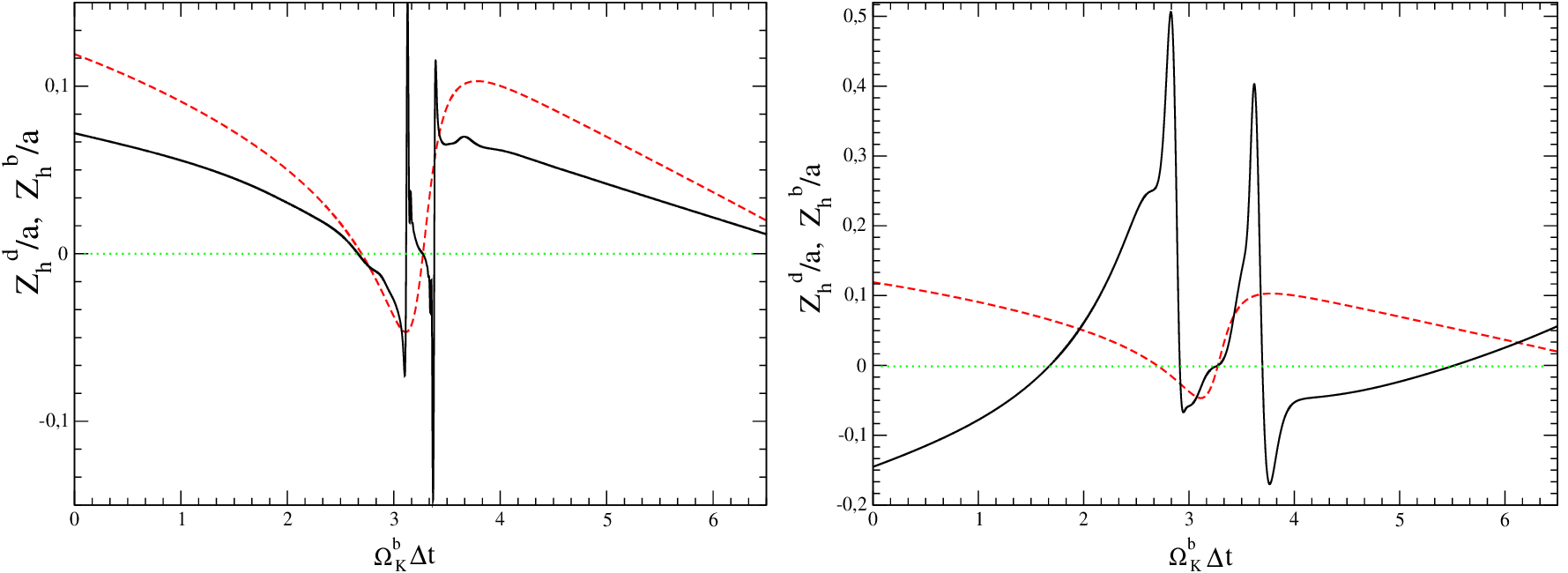}
\vspace{1cm}
\caption{Left panel. The solid curve shows the height of the disc above the equatorial 
plane of the primary black hole at the location of the perturber, $Z_h^d$, as a
function of time; the dashed curve shows the simultaneous height of the perturber, 
$Z_h^b$. 
Both quantities are expressed in units of $a$. The dotted line $Z_h=0$ shows 
the equatorial plane of the primary black hole.
The disc parameters are $\beta_b=0.1$ and $\chi=0.5$ 
and the disc evolution time is approximately $9\cdot 10^3{\Omega_K^b}^{-1}$.
Right panel. The same as the left panel, but for the retrograde 
rotation of the primary $\chi=-0.5$. Note that, for convenience, we shift the time origin in our figures
in such a way, that the orbit periastron corresponds to $\Omega^b_K\Delta t \approx 3$.}
\label{fig10}
\end{center}
\end{figure}

\begin{figure}
\begin{center}
\includegraphics[width=8.0cm,
angle=0]{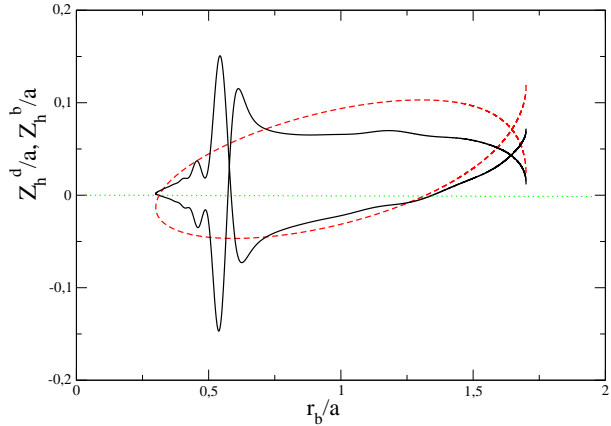}
\vspace{1cm}
\caption{Same as on the left panel in Fig.\ref{fig10}, but
the distance between the black holes, $r_b$ is used instead of time.}
\label{fig10a}
\end{center}
\end{figure}

\begin{figure}
\begin{center}
\includegraphics[width=16.0cm,
angle=0]{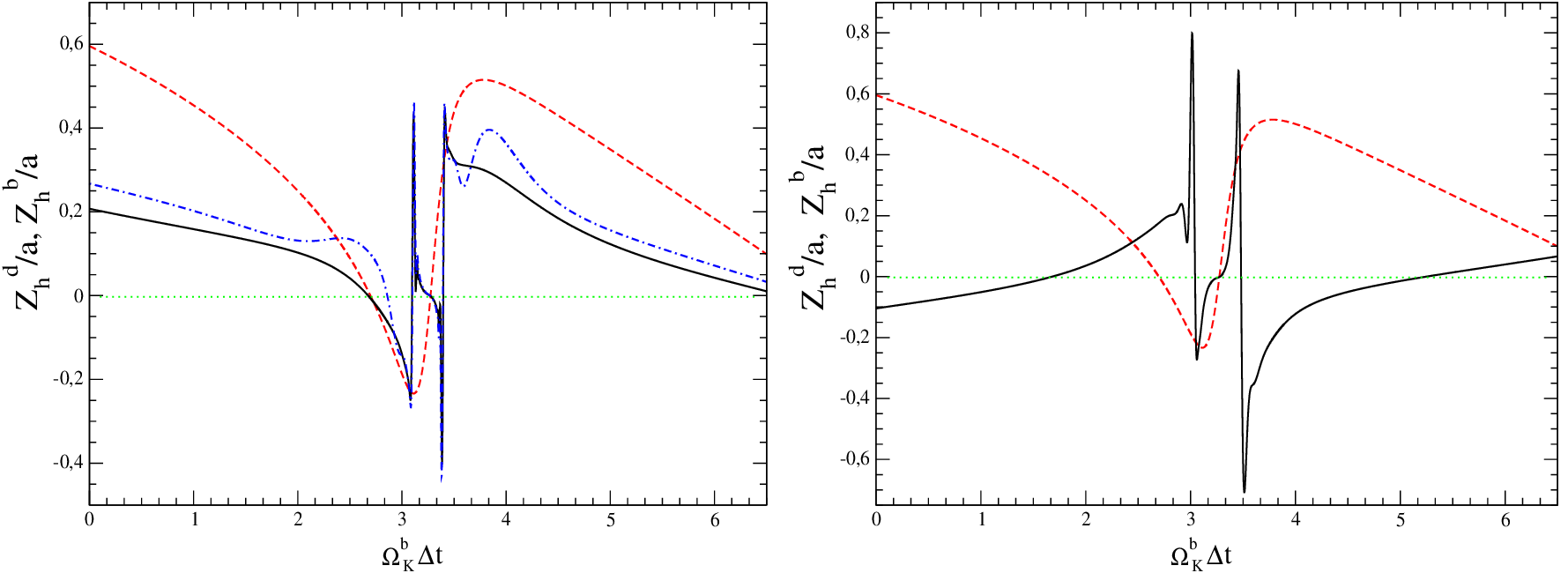}
\vspace{1cm}
\caption{The same as Fig.\ref{fig10}, but for the case $\beta_b=0.5$. Additionally, 
here we plot $Z^{d}_{d}$ corresponding 
to the model of disc evolution with $\Omega_2$ artificially set to zero by the dot dashed curve on the
left panel.}
\label{fig11}
\end{center}
\end{figure}

\begin{figure}
\begin{center}
\includegraphics[width=16.0cm,
angle=0]{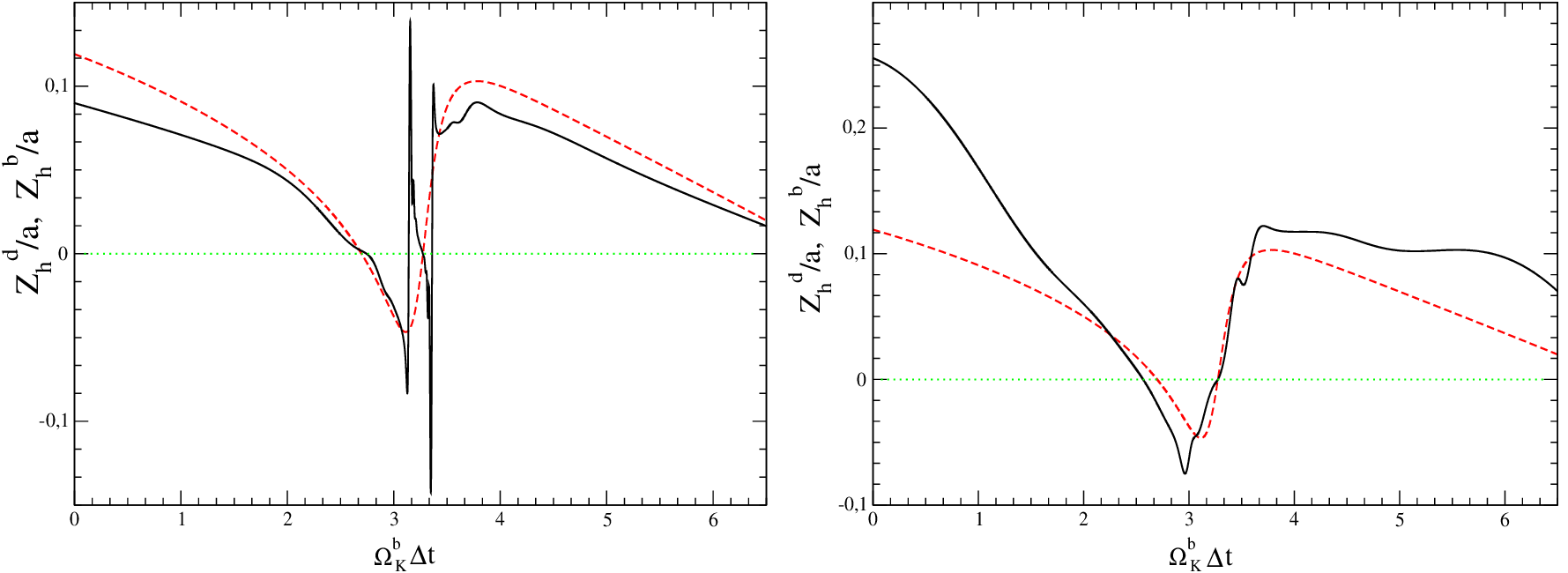}
\vspace{1cm}
\caption{The same as Fig. \ref{fig10}, but for smaller $|\chi|=0.25$.}
\label{fig12}
\end{center}
\end{figure}

\begin{figure}
\begin{center}
\includegraphics[width=16.0cm,
angle=0]{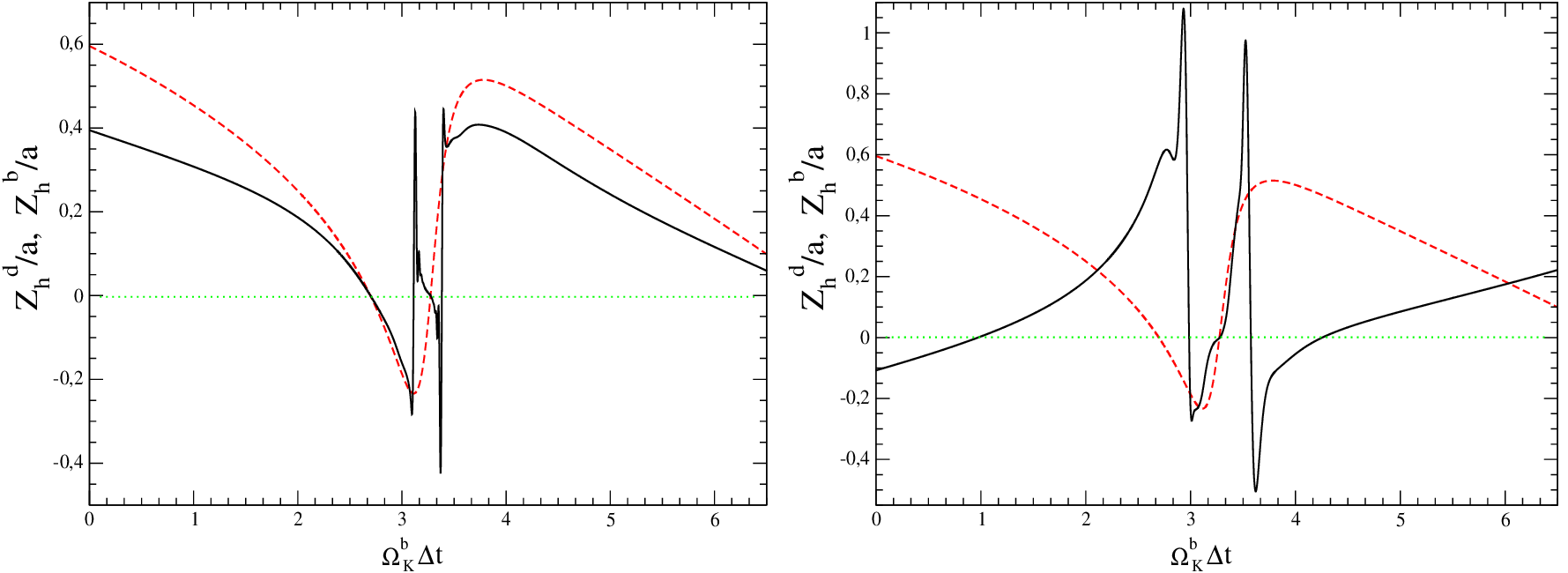}
\vspace{1cm}
\caption{The same as Fig. \ref{fig12}, but for the case $\beta_b=0.5$.}
\label{fig13}
\end{center}
\end{figure}

$Z_h^d$ and $Z_h^b$ are shown for a particular orbital period in 
Figs. (\ref{fig10}-\ref{fig13}). 
The instant values of $\Psi_b$ and $\gamma_b$ are obtained according to 
the uniform apsidal and nodal precessions of the binary orbit 
with the frequencies $\Omega_E$ and 
$\Omega^b_{LT}$, respectively, assuming that
$\Psi_b(t=0)=\pi/4$ and $\gamma_b(t=0)=0$. For our orbital parameters,
it gives the minimal time interval of the order of one year between the successive
intersections of the disc in the flat
disc case, which is broadly consistent with the PM model of OJ 287. Since we are going to discuss only
the principal importance of the disc tilt and twist, this appears to be sufficient for our study.   

\bigskip
\bigskip

\bigskip
\bigskip

\bigskip
\bigskip

\bigskip
\bigskip

\bigskip
\bigskip

In Figs. \ref{fig10} we show the dependencies of $Z_h^b$ and $Z_h^d$ on time during one orbital period,
for the disc model with $\beta_b=0.1$ and $|\chi|=0.5$ evolved for a sufficiently long time to the
quasi-stationary state. We see that both the number of crossings and the time passed between particular
crossings of the disc differ very significantly from the flat disc case. Formally, we have six crossings
of the disc instead of two in the flat case. Note, however, that certain crossings happen very close
to each other, as e.g. the ones close to $\Omega^b_K\Delta t \approx 3$ in the prograde case, these particular
pairs could merge if we take into account e.g. the finite thickness of the disc or the fact that 
the perturber is expected to expel the disc material within its accretion radius from the orbit. We further
illustrate this case in Fig. \ref{fig10a} showing the evolution of $Z_h^b$ and $Z_h^d$ on the plane $(r_h,Z_h)$,
where $r_h$ is the distance from the primary. One can see from this figure that the section
of this disc $(r_h,Z^d_h)$ has rather non-trivial from, with a 'butterfly-like' feature close to $r_h/a=0.5$.
This feature is explained using the results of Section \ref{analyt}. From these results it follows that
close to the resonance radius $r_r$ the disc's nodal angle $\gamma$ rotates by $\pi$. The same effect 
is also observed in our numerical calculations. This rotation leads to a very sharp variation of the
disc height along the orbit when the perturber's distance is close to $r_r$.  

We confirm that our main conclusion that the number of disc's crossings per period is significantly
modified is robust in Figs. \ref{fig11}-\ref{fig13}. In Fig. \ref{fig11} we illustrate the case with the same
value of black hole rotational parameter $|\chi|=0.5$, but for the orbit with $\beta_b=0.5$, while
in Fig. \ref{fig12} and \ref{fig13} we show the results for the case  $|\chi|=0.25$, for $\beta_b=0.1$
and $\beta_b=0.5$, respectively. One can see that the qualitative situation remains the same, although
the numbers of crossings and the crossing times differ from case to case. Additionally, on the left 
panel of Fig. \ref{fig11} we show the case of a disc model with the same parameters, but with $\Omega_2$
set artificially to zero. This corresponds to averaging of our disc model over a time period order
of $\Omega_{E}^{-1}$. It is seen again that that the situation remains very similar to the general case.   

\begin{figure}
\begin{center}
\includegraphics[width=16.0cm,
angle=0]{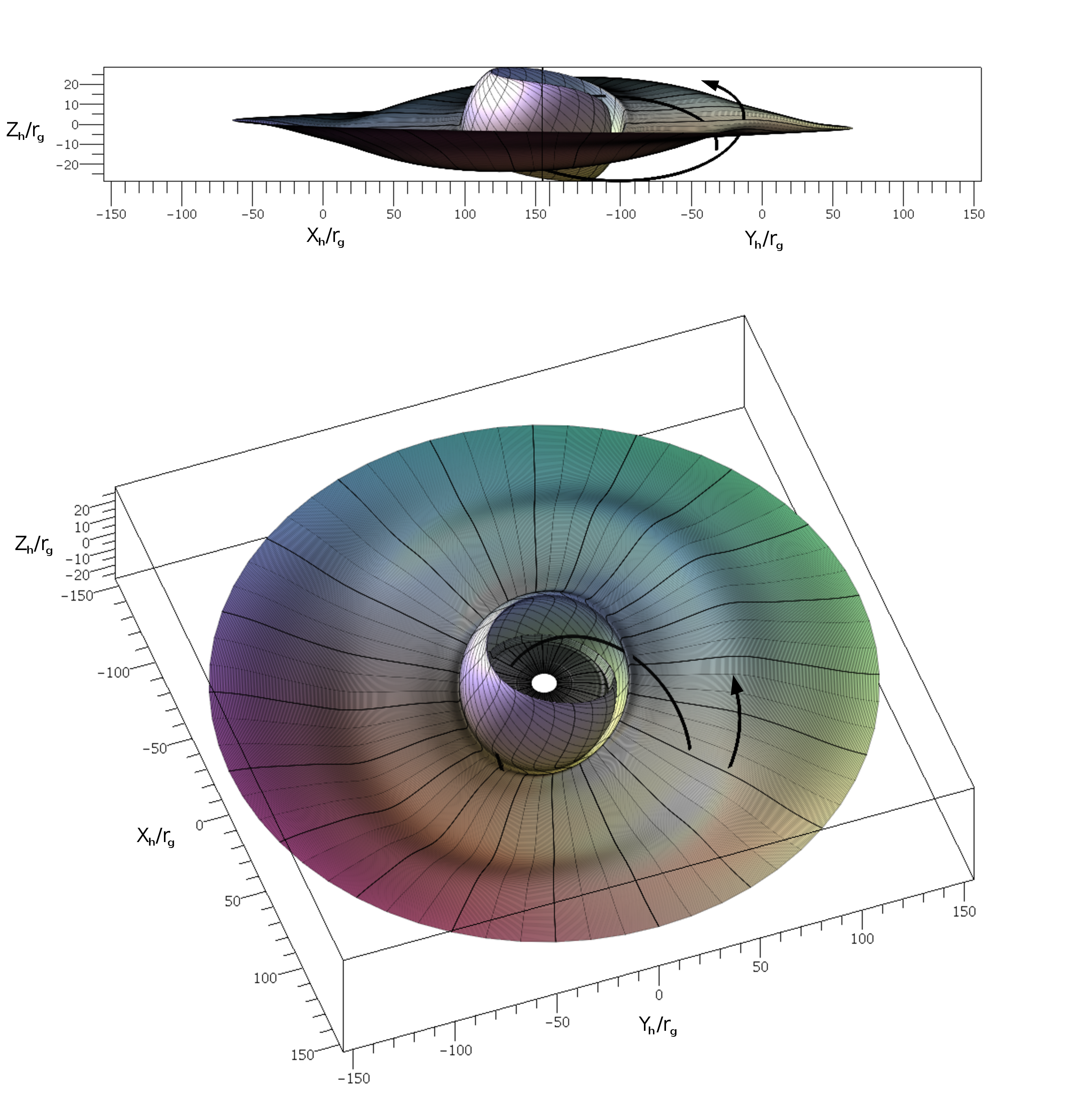}
\vspace{1cm}
\caption{A three-dimensional image of disc and its intersections with the binary orbit. The
trajectory of the secondary is shown by the black line with an arrow indicating the direction of the binary orbit. 
The disc parameters are $\beta_b=0.5$ and $\chi=0.5$ 
and the disc evolution time is close to that used in Figs. \ref{fig10}-\ref{fig13}.
Note that for this particular orbit its apsidal and nodal lines are close to each other.
Top panel. The observer's line of sight is in the equatorial plane of the primary
while it has an angle $-45$ degrees with respect to $X_h$-axis. 
Bottom panel. The observer's line of sight is inclined by $45$ degrees with 
respect to the equatorial plane of the primary while its projection onto this 
plane has an angle $-20$ degrees with respect to $X_h$-axis.
}
\label{fig_disc_3D}
\end{center}
\end{figure}

Finally, in Fig. \ref{fig_disc_3D} we show a three-dimensional picture of the disc, the perturber's orbit and
their intersections. One can clearly see a 'spiral rim' at radii close to the resonance radius in this Fig. 

\bigskip
\bigskip

\section{Conclusions and Discussion}
\label{conc}

In this Paper we considered effects caused by interaction of an eccentric inclined binary black hole with a very thin ($h/r \sim 10^{-3}$) accretion disc. The orbital parameters of the binary were assumed to be the same as in the so-called 'precessing massive' (PM) model of OJ 287 (see e.g. \cite{V23} and references therein). Namely, it was assumed that the binary consists of two black holes
with masses $M\approx 2\cdot 10^{10}M_{\odot}$ and $m\approx 10^{8}M_{\odot}$ on a very eccentric orbit with eccentricity $e=0.7$ 
and the semi-major axis $a\approx 60{GM\over c^2}$ corresponding to the orbital period in the observer's frame $\approx 12yr$. 
A distinctive feature of such a system is that its evolution time due to the emission of gravitational waves $T_{GW}$ is smaller 
than the disc relaxation time $T_{\nu}$, which suggests that the disc is unable to react to the presence of the binary by e.g.
opening a gap inside its orbit. Moreover, when the disc viscosity parameter $\alpha ~ > 0.01$, the evolution time is smaller
than the time $T_{al}$ associated with  alignment of the disc
to a quasi-stationary twisted configuration. It is this case that has been studied in this Paper in some
detail.

We calculated the torque 
 exerted on the disc 
by an inclined binary, assuming that the mass ratio $q$ is small and 
one can use a perturbation theory. Technically, we double averaged the expression for the gravitational force responsible for
this torque over the binary period and over a period of a particular disc element on a slightly perturbed circular orbit.
The same procedure is used to obtain a correction to the apsidal precession rate of the disc elements. We assumed that 
the double averaging procedure can be used to describe the evolution of the disc over characteristic time exceeding 
the binary orbital period and typical orbital periods of the disc elements. It was also assumed that the perturbing component
does not produce torques on the disc inside its 'accretion radius', since gas inside this radius is likely to be removed
from the system. In general, the corresponding torque and
the apsidal precession rate are obtained numerically, but simple asymptotic expressions can be found for the disc's elements
with  radii much smaller or much larger than $a$. We have checked that these expressions agree with similar expressions
obtained in other studies.

We used the torque term and the apsidal precession rate together with the standard expressions for these quantities induced by a relatively slowly rotating central black hole (primary) to calculate numerically the evolution of the disc tilt and twist from the flat configuration initially aligned with the equatorial plane of the primary. We set $\alpha=0.1$ for our runs
and used four values of the primary rotational parameter $\chi=\pm 0.25$ and $\chi=\pm 0.5$, where the plus (minus)
sign corresponds to prograde (retrograde ) rotation of the primary with respect to the direction of motion of the disc's elements.
We found that in all cases the disc relaxes to a non-trivial quasi-stationary configuration in the frame rotating with the Lense-Thirring frequency of the orbit $\Omega_{LT}^b$. The corresponding relaxation time of the order of the corresponding period $P_{LT}^b=2\pi/\Omega_{LT}^b$  is much smaller than $T_{GW}$. After the period of relaxation, the disc's inclination angle $\beta$ is almost 
stationary at scales order of $a$.
 It is shown that $\beta$  significantly varies inside the orbit. 
When $\alpha=0.1$, its value in the vicinity of its periastron is much smaller than the
orbital inclination angle $\beta_b$. At radii  between the periastron and apoastron of the orbit, $\beta$ becomes a
few times larger than $\beta_b$ while approaching 
back to $\beta_b$ in the vicinity of apoastron. 

{  Our preliminary analysis of the conditions of the crossing times made in Section \ref{inter}
shows that there are, in general, more intersections of the perturber with the twisted disc per
one orbital period that in the flat disc case. This happens in part due to the large disc's
tilt at scales of the order of the perturber's orbit, and, in part, due to a large disc's twist
near 'resonance' radius $r_r$, where the precession frequency of a disc element both due to the actions of the binary and the primary coincides with  $\Omega_{LT}^b$. Close to this resonance the disc's nodal angle
$\gamma$ rotates by $\pi$. Clearly, our results must be taken into account in the modelling of OJ 287 and
similar sources.}  

{  In Section \ref{analyt1} we proposed an elementary analytic model for the distribution of the disc inclination over $r$, which agrees with the numerical results surprisingly well apart from scales close to  $r_r$. A more
accurate, albeit more involved treatment was considered in Section \ref{analyt2}. It describes the quasi-stationary shape
of the disc in the frame precessing with the Lense-Thirring frequency of the binary at all radii, including the ones
close to the resonance. We compared the analytic and numerical results in Section \ref{comp}, and showed that
those are in good agreement. It is important to stress that one of the reasons of the agreement is that the term proportional to $\Omega_2$ in (\ref{e17c}) is not very important. It is proportional
to $e^{2i\Psi_b}$, and $\Psi_b$ evolves on a relatively fast timescale of Einstein's apsidal precession. However, the
disc tilt and twist mainly evolve on a longer timescale order of ${(\Omega^{b}_{LT})}^{-1}$, so that a contribution
of the term proportional to  $\Omega_2$ is averaged out.  Therefore, we can assume that 
the timescale of averaging the potential perturbations in Section \ref{calc} can be shorter than, but of the order
of ${(\Omega^{b}_{LT})}^{-1}$.    

Our analytic results can be helpful in calculating some approximate shapes of the disc without the use of numerical integration of the equations for
the disc's tilt and twist for other parameters of the system. For that, one can numerically evaluate the
distribution of
$\Omega_1$ using the results obtained in Section \ref{nod} for any desired parameters of the binary and the disc
and use results of Section \ref{analyt} to find the disc shape. This approach can, for example, 
be used in an accurate modelling of light curves of OJ 287 in the framework of the PM model, {  allowing for the possibility of multiple disc crossings per  orbital period found in this work}.
Of course, our approach is valid only for systems with sufficiently large values of $\alpha \gg \delta$.
   
In this Paper we did not make attempts to compare our model with observations. This requires a detailed survey of the parameter space which deserves a separate study with careful consideration of a lot of other factors influencing the determination of the light curve
(see \cite{Dey} and \cite{ZM} for a state-of-the-art modelling of the timing of OJ287 flares). Note, however, that it seems to be 
clear that taking into account of our results may lead to a significant revision of the parameters of the system in the framework
of PM model. Alternatively, it can  provide arguments in favour of other models, such as the models based on the jet activity (e.g. \cite{Br}) or, for example, the models of eccentric binary black holes of comparable masses immersed in a circumbinary disc, e.g. \cite{H1} and \cite{H2}.}

The numerical calculations should also be extended to smaller values of $\alpha$. As we briefly discussed in the text, smaller
values of $\alpha$ turn out to be demanding from the numerical point of view, and, therefore, numerical simulations of
the discs with lower viscosity are left for a future work. {  It is interesting to mention that the low viscosity discs can, in principle, have a qualitatively different behaviour. Namely, the presence of the resonance discussed in Section \ref{analyt} suggests a possibility of launching warp waves near the resonance by a mechanism similar to the well-known \cite{GT} mechanism of launching density waves near Lindblad resonances. This problem is beyond the scope of this work and is left for the future.}

Of course, our model should be checked with either SPH or grid based three-dimensional simulations. For that, one does not need to consider very thin discs or very long simulation times. It would suffice to check whether our expression for the binary torque
term is valid considering thicker discs on time scales of several $P_{LT}^b$.

{  Finally, it is interesting to find out how systems with the assumed parameters of OJ 287 can, in principle, be formed?
In this connection we would like to propose the following
scenario. It is known that outside the radius of influence of the more massive black hole the effects of dynamical friction could lead to 
the formation of a quite elongated orbit of the perturber under certain conditions (see e.g. \cite{PR} for analytic results and \cite{Q} for
numerical experiments). Assuming that these conditions are fulfilled, a characteristic time of a significant change in orbital angular momentum
due to gravitational scattering of stars of the central cluster is smaller than that of a significant change in the orbital size. 
In the course of  
the orbital evolution, the angular momentum could acquire very small values corresponding to impact parameters order of hundreds of gravitational radii of the primary black hole. In this situation, the process of direct capture of the perturber onto a bound orbit due to emission of gravitational waves is possible and a simple estimate shows that this effect may be realistic for systems with parameters appropriate to those required in galactic centres
harbouring black holes as massive as $10^{10}M_{\odot}$. To the best of our knowledge, such a scenario has not yet been considered. Its analysis is left for a future work.}

\section*{Acknowledgments}

VVZ acknowledges the support of Roscosmos.  We are
grateful to P. C. Fragile, K. Hayasaki, J. C. B. Papaloizou and E. A. Vasiliev for useful comments, K. Zhuravleva for
help and the referee for
a very helpful report.

\section{DATA AVAILABILITY}

There are no new data associated with this article.

\begin{appendix}

\end{appendix}

\label{lastpage}


\begin{thebibliography}{33}

\bibitem[\protect\citeauthoryear{Abod et al}{2022}]{Ab}
Abod, C. P., Chen, C.,  Smallwood, J.,  Rabago, I,  Martin, R. G.,  Lubow, S. H., 2022
MNRAS, 517, 732
\bibitem[\protect\citeauthoryear{Aly et al}{2015}]{Aly}
Aly, H., Dehnen, W., Nixon, C., King, A., 2015, MNRAS, 449, 65
MNRAS, 517, 732
\bibitem[\protect\citeauthoryear{Artymowicz}{1994}]{A} Artymowicz, P., 1994, ApJ, 423, 581
\bibitem[\protect\citeauthoryear{Artymowicz \& Lubow}{1994}]{AL}  Artymowicz, P., Lubow, S.
H., 1994, ApJ, 421, 651
\bibitem[\protect\citeauthoryear{Britzen et al}{2023}]{Br} 
Britzen, S., Zajacek, M., Gopal-Krishna, Fendt, C.,  Kun, E.,  Jaron, F.,  Sillanpää, A.,  Eckart, A., 2023,
ApJ, l, 951, 106   
\bibitem[\protect\citeauthoryear{Demianski \& Ivanov}{1997}]{DI} Demianski, M., Ivanov, P. B.,
1997, A$\&$A, 324, 829
\bibitem[\protect\citeauthoryear{Dey et al}{2018}]{Dey} Dey, L. at al, 2018, ApJ, 866, 11 
\bibitem[\protect\citeauthoryear{Dogan et al }{2015}]{DNKP} Dogan, S., Nixon, C.,  King, A., Price, D.J.,
2015, MNRAS, 449, 1251
\bibitem[\protect\citeauthoryear{Dotti, Sesana \& Decarli }{2012}]{Dot} Dotti, M., Sesana, A., Decarli, R., 
2021, Advances in Astronomy, 2012, 940568
\bibitem[\protect\citeauthoryear{Eracleous et al}{2012}]{Err}Eracleous, M.,  Boroson, T. A., Halpern, J. P.,  Liu, J.,
2012, ApJ Suppl, 201, 23
\bibitem[\protect\citeauthoryear{Ferrarese \& Ford}{2005}]{Fer} Ferrarese, L., Ford, H., 2005, Space Science Reviews, 116, 523
\bibitem[\protect\citeauthoryear{Graham et al}{2015}]{Grah} Graham, M. J.,  Djorgovski, S. G.,  Stern, D.,  Drake, A. J.,  Mahabal, A. A.,
Donalek, C., Glikman, E.,  Larson, S.,  Christensen, E., 2015, MNRAS, 
453, 1562
\bibitem[\protect\citeauthoryear{Goldreich \& Tremaine}{1979}]{GT}
Goldreich, P., Tremaine, S., 1979, ApJ,  233, 857 
\bibitem[\protect\citeauthoryear{Gurkan \& Hopman }{2007}]{GH}
Gurkan, M. A.,  Hopman, C., MNRAS, 379, 1083
\bibitem[\protect\citeauthoryear{Hayasaki, Mineshige \& Sudou }{2007}]{H1}
Hayasaki, K., Mineshige, S., Sudou, H., 2007, Publ. Astron. Soc. Japan, 59, 427 
\bibitem[\protect\citeauthoryear{Hayasaki, Mineshige \& Ho}{2008}]{H2} 
Hayasaki, K.,  Mineshige, S., Ho, L. C., 2008, ApJ, 682, 1134
\bibitem[\protect\citeauthoryear{Hayasaki, Saito \& Mineshige }{2013}]{HSM}
Hayasaki, K., Saito, H.,  Mineshige, S., 2013, Publ. Astron. Soc. Japan, 65, 86
\bibitem[\protect\citeauthoryear{Hayasaki et al}{2015}]{HSOJZN}
Hayasaki, K.,  Sohn, B. W., Okazaki, A. T.,  Jung, T.,  Zhao, G.,  Naito, T., 2015,
JCAP, 07, 005
\bibitem[\protect\citeauthoryear{Hourigan \& Ward}{1984}]{HW} Hourigan, K., Ward, W. R.,
1984, Icarus, 60, 29
\bibitem[\protect\citeauthoryear{Ivanov \& Illarionov}{1997}]{II} 
Ivanov, P. B., Illarionov, A. F. 1997, MNRAS, 285, 394
\bibitem[\protect\citeauthoryear{Ivanov, Igumenshchev  \& Novikov}{1998}]{IIN} 
Ivanov, P. B., Igumenshchev, I. V., Novikov, I. D., 1998, ApJ , 507, 131
\bibitem[\protect\citeauthoryear{Ivanov, Papaloizou  \& Polnarev}{1999}]{IPP} Ivanov, P. B.,  
Papaloizou, J. C. B., Polnarev, A. G., 1999,  MNRAS, 307, 79
\bibitem[\protect\citeauthoryear{Komossa et al}{2023}]{Komossa} Komossa, S.,  Grupe, D., Kraus, A., Gurwell, M. A.,  Haiman, Z.,  Liu, F. K.,
Tchekhovskoy, A.,  Gallo, L. C.,  Berton, M.,  Blandford, R.,  Gómez, J. L.,  Gonzalez, A. G.,
2023, MNRAS, 522, L84
\bibitem[\protect\citeauthoryear{Komossa}{2006}]{Kom1} Komossa, S., 2006, Memorie della Società Astronomica Italiana, 77, 733 
\bibitem[\protect\citeauthoryear{Lin  \& Papaloizou}{1986}]{LinP} Lin, D. N. C., Papaloizou, 
J. C. B., 1986, ApJ, 309, 846
\bibitem[\protect\citeauthoryear{Larwood  \& Papaloizou}{1997}]{LP} 
Larwood, J. D., Papaloizou, J. C. B., 1997, MNRAS, 285, 288
\bibitem[\protect\citeauthoryear{Lehto  \& Valtonen}{1996}]{LV}Lehto, H. J., Valtonen, M. J., 1996
ApJ, 460, 207
\bibitem[\protect\citeauthoryear{Malik et al}{2015}]{MMMM} Malik, M, Meru, F., Mayer, L., 
Meyer, M., 2015, ApJ, 802, 56
\bibitem[\protect\citeauthoryear{ Malinovsky \& Mikheeva}{2023}]{amalin} Malinovsky, A. M., Mikheeva, E. V., Astronomy Reports, 67, 685
\bibitem[\protect\citeauthoryear{ Meritt}{2013}]{Meritt} Meritt, D., 2013, in Dynamics and Evolution of Galactic Nuclei. Princeton, NJ: Princeton University Press 
\bibitem[\protect\citeauthoryear{Morales \& Teixeira et al }{2014}]{TFZI}
Morales Teixeira D., Fragile P. C., Zhuravlev V. V., Ivanov P. B., 2014, ApJ,
796, 103
\bibitem[\protect\citeauthoryear{Novikov \& Thorne} {1973}]{NT} Novikov, I. D.,  Thorne, K. 1973, in Black Holes, ed. C. DeWitt $\&$ B. DeWitt
(New York: Gordon and Breach), 343 
\bibitem[\protect\citeauthoryear{Ogilvie} {1999}]{O1} Ogilvie, G. I., 1999, MNRAS, 304, 557
\bibitem[\protect\citeauthoryear{Ogilvie} {2006}]{O2} Ogilvie, G. I., 2006, MNRAS, 365, 977
\bibitem[\protect\citeauthoryear{Ogilvie \& Latter} {2013}]{LO} Ogilvie, G. I., Latter, H. N., 2013, MNRAS, 
433, 2403
\bibitem[\protect\citeauthoryear{Olver et al} {2010}]{Olv} Olver, F. W. J., Lozier, D., M., Boisvert, R. F., Clark, C. W., 2010, NIST Handbook of Mathematical Functions, Cambridge University Press
\bibitem[\protect\citeauthoryear{Papaloizou \& Pringle} {1983}]{PP83} 
Papaloizou J. C. B., Pringle J. E., 1983, MNRAS, 202, 1181
\bibitem[\protect\citeauthoryear{Peters} {1964}]{P64} Peters, P. C., 1964,  
Physical Review, 136, 1224
\bibitem[\protect\citeauthoryear{Petterson} {1978}]{Pet} Petterson, J. A., ApJ 226, 253, 1978
\bibitem[\protect\citeauthoryear{Polnarev \&  Rees}{1994}]{PR} Polnarev, A. G.,  Rees, M. J.,
1994, A$\&$A, 283, 301 
\bibitem[\protect\citeauthoryear{ Quinlan \& Hernquist}{1997}]{Q} Quinlan, G. D.,  Hernquist, L., 1997, New Astron., 2, 533
\bibitem[\protect\citeauthoryear{Rauch \& Tremaine}{1996}]{RT} Rauch K. P., Tremaine S., 1996, New Astron., 1, 149
\bibitem[\protect\citeauthoryear{Shakura \& Sunyaev}{1973}]{SS} Shakura, N. I., Sunyaev, R. A., 1973, A$\&$A, 24, 337
\bibitem[\protect\citeauthoryear{Sillanpaa et al}{1988}]{Sill} Sillanpaa, A., Haarala, S.,  Valtonen, M. J.,  Sundelius, B., Byrd, G. G.,
1988, ApJ, 325, 628
\bibitem[\protect\citeauthoryear{Valtonen et al}{2008}]{V08}
Valtonen, M. J.,  Lehto, H. J.,  Nilsson, K.,  Heidt, J.,  Takalo, L. O.,  Sillanpaa, A.,  Villforth, C.,  
Kidger, M.,  Poyner, G.,  Pursimo, T.,  Zola, S.,  Wu, J. -H.,  Zhou, X., Sadakane, K., 
Drozdz, M.,  Koziel, D.,  Marchev, D.,  Ogloza, W.,  Porowski, C.,  Siwak, M. 2008, Nature,   
Nature, 452, 851
\bibitem[\protect\citeauthoryear{Valtonen et al}{2023}]{V23} Valtonen, M. J., Zola, S., Gopakumar, A.,
Lahteenmaki, A., Tornikoski, M.,
Dey, L. , Gupta, A. C., Pursimo, T., Knudstrup, E., Gomez, J. L.,
Hudec, R., Jelinek, M.,  Strobl, J.,
Berdyugin, A. V., Ciprini, S., 
Reichart, D. E., Kouprianov, V. V., Matsumoto, K.,  Drozdz, M.,
Mugrauer, M.,  Sadun, A.,  Zejmo, M., Sillanpaa, A., Lehto, H. J., Nilsson, K.,
Imazawa, R., Uemura M., 2023, MNRAS 521, 6143
\bibitem[\protect\citeauthoryear{ Xiang-Gruess \& Papaloizou}{2014}]{XP} Xiang-Gruess, M., Papaloizou,
J. C. B., 2014, MNRAS, 440, 1179  
\bibitem[\protect\citeauthoryear{Young et al}{2023}]{Y} Young, A. K., Stevenson, S., Nixon, C. J.,  Rice, K.,  MNRAS, Advance Access
\bibitem[\protect\citeauthoryear{Zanazzi \& Lai}{2018}]{ZL} Zanazzi, J. J., Lai, D.,
2018, MNRAS 473, 603
\bibitem[\protect\citeauthoryear{Zhuravlev \& Ivanov}{2011}]{ZI}
Zhuravlev, V. V., Ivanov, P. B., 2011, MNRAS, 415, 2122
\bibitem[\protect\citeauthoryear{Zhuravlev et al}{2014}]{ZIFT} 
Zhuravlev V. V., Ivanov P. B., Fragile P. C., Teixeira D. M., 2014, ApJ, 796, 104
\bibitem[\protect\citeauthoryear{Zwick \& Mayer}{2023}]{ZM} 
Zwick, L., Mayer, L., 2023, MNRAS, 526, 2754
\end{thebibliography}
\end{document}